\documentclass[%
preprint,
 amsmath,amssymb,
 aps,
]{revtex4-2}

\usepackage{graphicx}
\usepackage{dcolumn}
\usepackage{bm}
\usepackage{hyperref}
\usepackage[OT2,T1]{fontenc}
\usepackage{ae} 
\usepackage{aecompl}
\DeclareSymbolFont{cyrletters}{OT2}{wncyr}{m}{n}
\DeclareMathSymbol{\Sha}{\mathalpha}{cyrletters}{"58}

\begin{document}


\title{Rational Design Protocols for Size-Based Particle Sorting Microdevices Using Symmetry-Induced Cyclical Dynamics}

\author{Arnaldo Rodriguez-Gonzalez}
\email{ajr295@cornell.edu}
\affiliation{Field of Theoretical \& Applied Mechanics, Sibley School of Mechanical and Aerospace Engineering, Cornell University, Ithaca NY 14853}

\author{Jason P. Gleghorn}
\affiliation{Department of Biomedical Engineering, University of Delaware, Newark, DE 19716}%

\author{Brian J. Kirby}%
\email{bk88@cornell.edu}
\affiliation{Sibley School of Mechanical and Aerospace Engineering, Cornell University, Ithaca NY 14853}
\affiliation{Department of Medicine, Division of Hematology and Medical Oncology, Weill-Cornell Medicine, New York NY 10065}

\date{\today}

\begin{abstract}
In this paper, we describe the unification and extension of multiple kinematic theories on the advection of colloidal particles through periodic obstacle lattices of arbitrary geometry and infinitesimally small obstacle size. We focus specifically on the particle displacement lateral to the flow direction (termed deterministic lateral displacement or DLD) and the particle-obstacle interaction frequency, and develop novel methods for describing these as a function of particle size and lattice parameters for arbitrary lattice geometries for the first time in the literature. We then demonstrate design algorithms for microfluidic devices consisting of chained obstacle lattices of this type that approximate any lateral displacement function of size to arbitrary accuracy with respect to multiple optimization metrics, prove their validity mathematically, and compare the generated results favorably to designs in the literature with respect to metrics such as accuracy, device size, and complexity.
\end{abstract}

\maketitle


\section{Introduction}
\begin{figure}[h]
\centering
\includegraphics{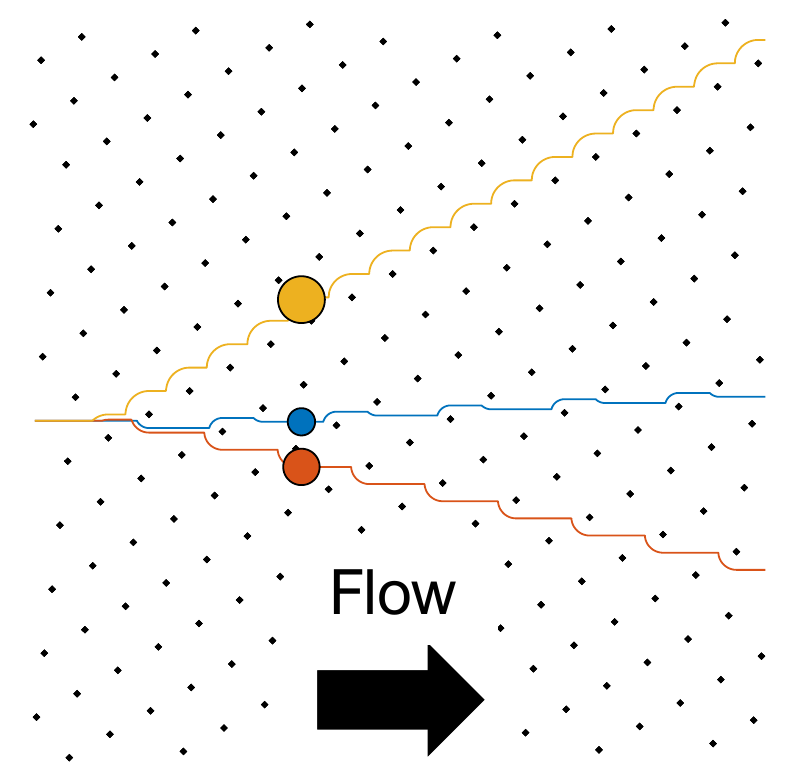}
\caption{\label{fig:devicevisuals} A visual representation of round particles advecting through a channel patterned with a lattice of obstacles and their trajectories. The particles, upon interaction with the obstacles, displace lateral to the direction of the flow and periodically collide with the obstacles in a way that is strongly dependent on their size.}
\end{figure}

The advection of particles through obstacle lattices, as illustrated in Figure \ref{fig:devicevisuals}, has been presented in the literature in multiple contexts, focusing on two particular aspects; interaction-induced particle displacement lateral to the flow direction (first reported and termed "deterministic lateral displacement" by \cite{Huang987}) and spatial particle--obstacle collision frequency (utilized in methods such as geometrically-enhanced differential immunocapture by \cite{10.1371/journal.pone.0035976}). This process of lateral particle displacement has seen a varied scope of applications to particle and cell sorting in microfluidic devices, compiled in detail in \cite{C4LC00939H}, and as a result has been described in the literature in terms of a multiplicity of descriptors specific to particular lattice geometries. Spatial collision frequencies and its dependence on size and obstacle array parameters have also been studied for specific lattice geometries in work such as \cite{PhysRevE.88.032136}. Methods to understand the general relationship between this displacement and particle-obstacle collision frequencies, however, have been largely unexplored. As a result, inverse design protocols for fabrication of these devices has not been systematically developed.

In this manuscript, we discuss a unified mathematical framework to describe particle trajectories in arbitrary lattices of infinitesimally small obstacles, and quantify key readouts such as lateral displacement and spatial collision frequency. We also discuss how these displacement \& collision readouts change as a function of the design parameters of the lattice, and describe a formal mathematical framework for these readouts in microdevices with sequences of chained arbitrary obstacle lattices of this type. We then describe an algorithmic solution for the inverse design problem of constructing such a device to approximate a given size-dependent lateral displacement function. We show how current representations in the literature in the limit of small obstacle size are simplifications or special cases of those derived from this model, clarify and interpret the scattered language present in the literature from previous work by multiple research groups, and compare microdevice designs constructed from our algorithm to other devices of this type in the literature.

\section{Particle Trajectories Through Obstacle Lattices}

Microdevices in the literature exploiting periodic particle--obstacle interaction generally consist of a microfluidic channel with a repeating pattern of "posts" or obstacles within it. Such devices have a straightforward phenomenological description of their operation; a steady stream of particles enters the device from an inlet, and these particles flow downstream through the device while interacting with these obstacles, causing the particles to displace laterally due to scattering-like deflections. Consequently, the lateral displacement of particles and the number of particle--obstacle interaction events are the principal readouts of interest, and the configuration and design of these microdevices is dependent on which of these readouts is considered to be of principal importance.

In order to understand the design of such devices and to aid the mathematical analyses of the particle dynamics that is to follow, we describe the geometry of these periodic obstacle patterns in a general way using lattices, implicitly assuming the obstacles are sufficiently small as to be approximated by such constructions, and follow by describing common experimental actualizations of these patterns.

\subsection{Obstacle Lattice Geometry\label{latticegeom}}

We define a 2-dimensional obstacle lattice using lattice vectors $\vec{l_a},\ \vec{l_b}$ representing the crystal basis of a given lattice structure. We can then define a transformation matrix $A$ that maps lattice coordinates to real space:
\begin{equation}
\begin{bmatrix} \vec{l_a} & \vec{l_b} \end{bmatrix}\begin{bmatrix}a \\ b \end{bmatrix} = \begin{bmatrix} A \end{bmatrix}\begin{bmatrix}a \\ b \end{bmatrix} = \begin{bmatrix}x \\ y \end{bmatrix}
\end{equation}

Here $a, b \in \mathbb{Z}$ are lattice coordinates that represent how many obstacles away an arbitrary obstacle is in the direction of the lattice vectors $\vec{l_a},\ \vec{l_b}$ respectively, whereas $x, y$ specify the location of an arbitrary obstacle in Cartesian coordinates using the origin obstacle as the origin and defining the x-axis parallel to the flow. Without loss of generality, we can define the lattice vector $\vec{l}_a$ as the lattice vector in the unit cell possessing the largest projection onto the streamwise axis $\hat{x}$, and the lattice vector $\vec{l}_b$ as the lattice vector in the unit cell with the largest projection onto the lateral axis $\hat{y}$.

Consequently, all the details of the structure of the lattice are contained in the transformation matrix $A$, and thus all specific design parameters for some obstacle lattice are encoded in its four elements. A given lattice, however, is not uniquely described by a set of lattice vectors; multiple combinations of lattice vectors give rise to the same lattice.

We also impose a constraint on the lattice geometry such that the Euclidean distance between every obstacle must be larger than the diameter of the largest particle that will flow through the lattice. This is done to prevent particle clogging and to ensure that particles interact with only one obstacle at a time. By defining quantities $||\vec{l_c}||_{2}, ||\vec{l_d}||_{2}$ as magnitudes of lattice vectors pointing in the direction of the nearest neighboring obstacles, this is enforced quantitatively by $||\vec{l_c}||_{2}, ||\vec{l_d}||_{2}, ||\vec{l_c} + \vec{l_d}||_{2} \geq 2r_{\mathrm{max}}$, where $||.||_{2}$ is the Euclidean norm and $r_{\mathrm{max}}$ is the largest particle flowing through the obstacle lattice.

\subsubsection{Example: Square Lattices\label{rotsq}}

\begin{figure}
\centering
\includegraphics{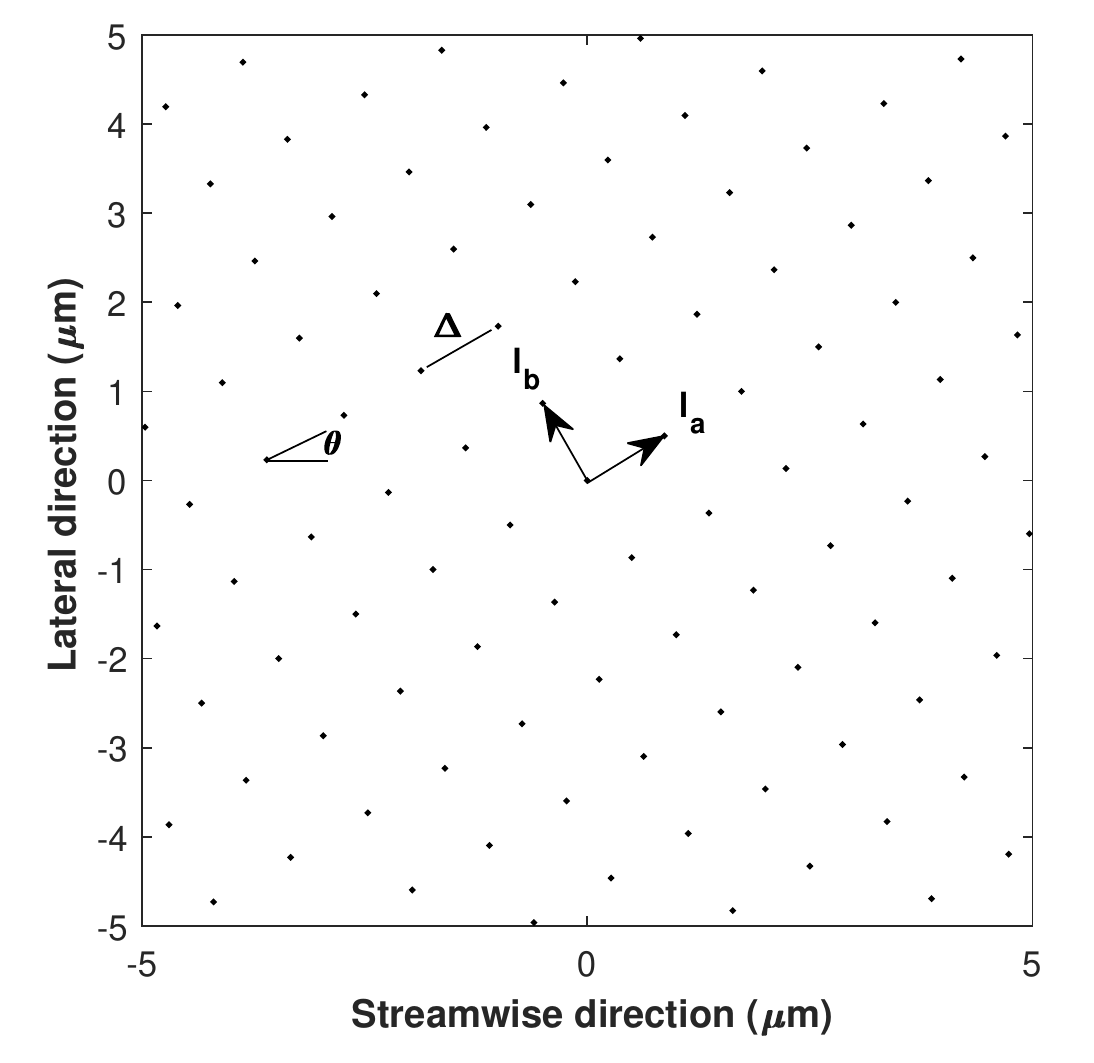}
\caption{\label{fig:square} Diagram of square lattice geometry with corresponding lattice basis vectors, $\theta = 30^{o}$ and $\Delta = 1 \ \mu\text{m}$.}
\end{figure}

For a 2D square lattice (tetragonal; symmetry group p4m) rotated counterclockwise at an arbitrary angle relative to the fluid flow, which can be characterized by an obstacle spacing $\Delta$ and a rotation angle $\theta$. These lattices are of the type considered in \cite{risbud_drazer_2013} \& \cite{PhysRevLett.92.130602}, which have been called ``rotated squares'' and which have the attractive property of isotropic fluid permeability. By selecting $l_a$ and $l_b$ as vectors of length $\Delta$ rotated by $\theta$ relative to the $x$ and $y$ axes, we find that

\begin{equation}
A_{\mathrm{s.l.}} = \begin{bmatrix} \Delta\cos\theta & -\Delta\sin\theta \\ \Delta\sin\theta & \Delta\cos\theta \end{bmatrix} = \Delta\begin{bmatrix} \cos\theta & -\sin\theta \\ \sin\theta & \cos\theta \end{bmatrix}
\end{equation}

which is simply a rotation matrix for a Cartesian coordinate system scaled by the obstacle spacing $\Delta$. Figure \ref{fig:square} shows a diagram of a square lattice and the lattice vectors representing the crystal basis we detail above.

\subsubsection{Example: Oblique Lattice}

Here we consider a 2D oblique lattice (monoclinic; symmetry group p2) of the type described in \cite{PhysRevE.88.032136} and \cite{Kim201706645}, which can be obtained by adding a vertical offset to each successive row in a non-rotated square lattice, and which have been termed ``row-shifted parallelograms''. For an oblique lattice whose rows are aligned normal to the flow, there are three key design parameters: the row spacing $\Gamma$, column spacing $\Lambda$, and offset $\Pi$. By selecting $\vec{l_a}$ as the vector pointing in the direction of the closest downstream obstacle in its row and $\vec{l_b}$ as the vector pointing to the next downwards obstacle in its column, we find that
\begin{equation}
A_{\mathrm{o.l.}} = \begin{bmatrix} \Lambda & 0 \\ \Pi & -\Gamma \end{bmatrix}
\end{equation}

\begin{figure}
\centering
\includegraphics{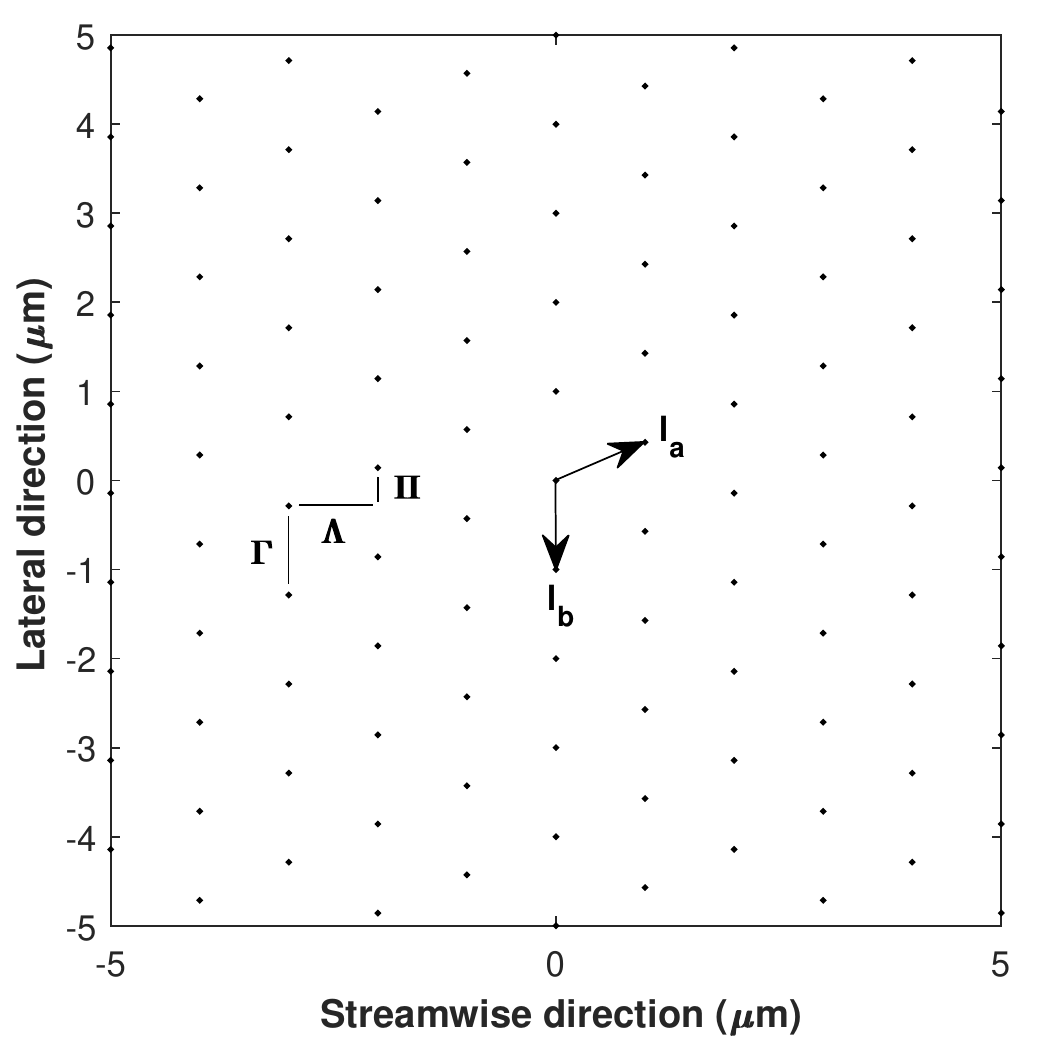}
\caption{\label{fig:oblique} Diagram of oblique lattice geometry with corresponding lattice basis vectors, $\Lambda = 1\ \mu\text{m}$, $\Pi = \frac{3}{7}\ \mu\text{m}$ and $\Gamma = 1 \ \mu\text{m}$. Note that the $\vec{l_a}$ lattice basis vector is defined to point to the nearest downstream lattice point.}
\end{figure}

See Figure \ref{fig:oblique} for a diagram of an oblique lattice and its corresponding lattice basis vectors as defined above.

\subsection{Particle Trajectories\label{traj}}

The particle trajectories in these obstacle lattices are dictated primarily by particle-obstacle collisions and only secondarily by fluid-mechanical phenomena. Hence in this manuscript, we consider a limit that highlights the primary role of lattice geometries on particle advection. Although these are gross simplifications, they reproduce most of the described behavior. Consequently, we consider flow through these obstacle lattices in the limit of Hele-Shaw flow, and model the lattice obstacles as infinitesimally small. As a result of modeling the obstacle lattices as point lattices, the obstacles perturb the fluid only in an infinitesimally small region. By also assuming the particles are rigid, spherical, and inertialess, their motion can be assumed to be uniform except when interacting with an obstacle, and the flow is uniform and Reynolds-number-independent.

We can model the trajectories of the particles by approximating the short-range nonlinear forces between a particle and an obstacle as a contact force. Hence, a particle interacting with an obstacle in this model will collide with an obstacle, displace laterally until it does no longer feels a contact force, and translate uniformly on its new streamline (shifted laterally by $\pm r$ from the position of the obstacle) until it collides with another obstacle in the lattice. This model allows us to treat each particle-obstacle interaction individually, without considering the effects of neighboring obstacles or particles. The trajectories of particles in this type of kinematic model are largely consistent with those observed in real systems\cite{PhysRevE.88.032136}, and can be readily extended to systems with obstacles of finite size (see Section \ref{discussion}).

\subsection{Symmetry-Induced Cyclical Dynamics}

As a result of both the periodic nature of the lattice and the size-dependent particle--obstacle interactions, the lateral displacement and spatial collision frequencies of particles advecting through an obstacle lattice are strongly dependent on their size. Previous theoretical models in conjunction with experimental results\cite{PhysRevLett.92.130602} \cite{PhysRevE.88.032136} \cite{PhysRevE.90.012302} describe these particles as eventually advecting in strictly periodic trajectories referred to as ``modes'', although the terminology associated with this description has varied between authors. Those focused on lateral displacement \cite{Huang987} have distinguished between modes with and without lateral displacement, whereas those focused on collisions \cite{PhysRevE.88.032136} have used language that described the nature of the collisions relative to rows. Here we show that our model generates and completely describes these periodic trajectories as a direct consequence of the the finite number of particle--obstacle interaction outcomes and the spatial periodicity of the obstacle lattice, and enforces particles to advect in these periodic trajectories after three interactions at most---a phenomenon we term "symmetry-induced cyclical dynamics".

\begin{figure}[h]
\centering
\includegraphics[width=0.9\textwidth]{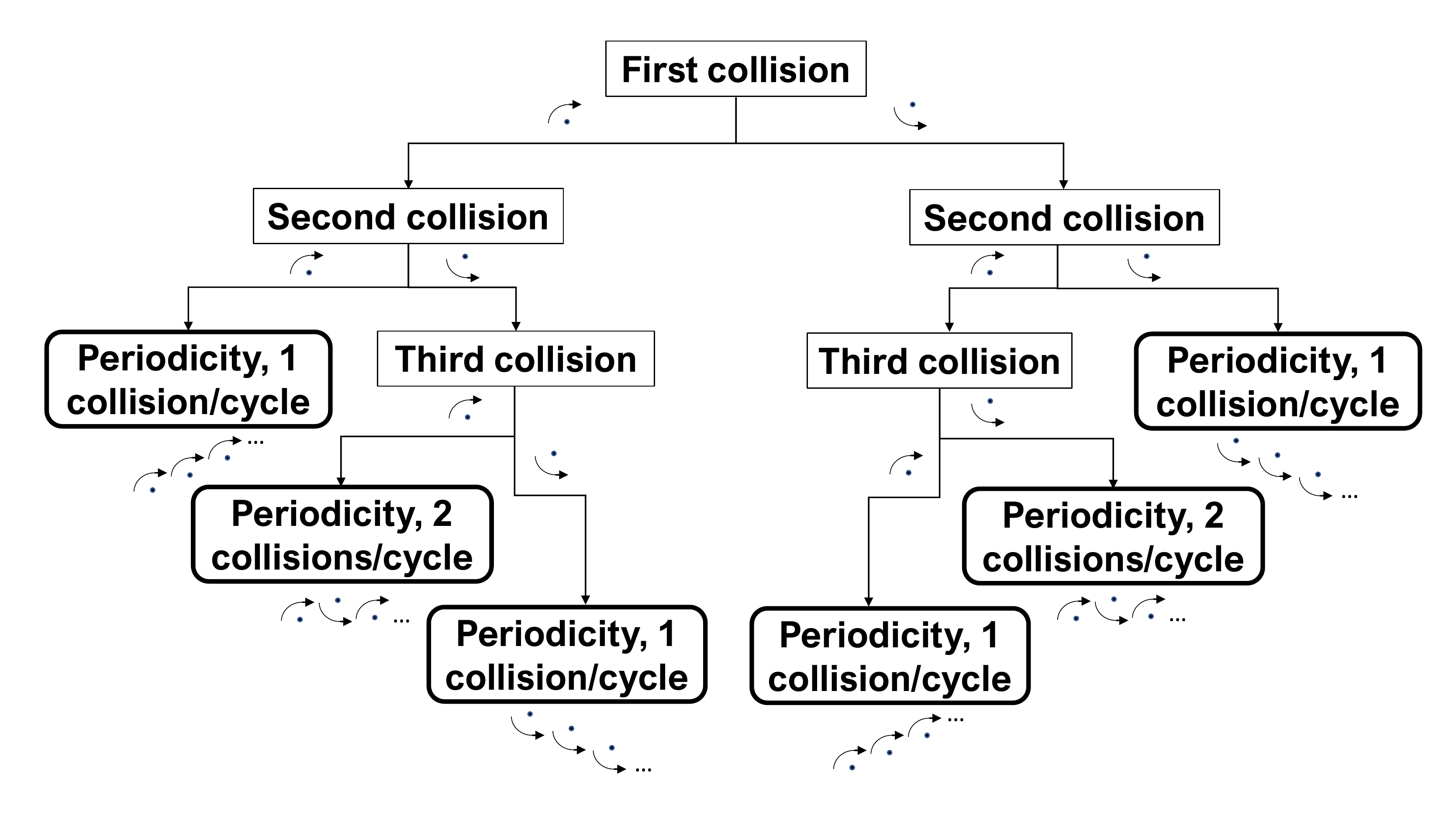}
\caption{\label{fig:flowchart} Decision tree representing the distinct particle-obstacle collision outcomes of a particle flowing downstream through an obstacle lattice, where the upwards or downwards arrow-dot symbols represent lateral displacements of magnitude $r$ from the location of the obstacle in the respective lateral direction. By the third collision, an outcome is always repeated and translational symmetry of the trajectory has been established. The resultant trajectories may only have either one or two collisions per cycle.}
\end{figure}

To illustrate this, consider a particle interacting sequentially with three obstacles in a periodic lattice. Because the only outcomes from a particle--obstacle interaction are a lateral displacement of either $+r$ or $-r$ from the location of the obstacle the particle interacted with, the particle will have spanned all of the possible outcome branches by the time of its third interaction, and will inevitably repeat an outcome it has previously sampled. Due to the translational symmetry of the lattice, the particle must then repeat the interactions that placed it on that outcome, generating a periodic trajectory in which the particle must either displace laterally in only one direction (which we label a "pure" trajectory) or must displace laterally in strictly alternating directions (which we label a "mixed" trajectory). See Figure \ref{fig:flowchart} for a diagram of this process.

More formally, the spatial periodicity of the particle--obstacle interactions induces a spatial periodicity on the particle trajectories; hence the term ''symmetry-induced cyclical dynamics". Consequently, the translational symmetry group of the particle trajectories must be a cyclic subgroup of the lattice group, and the generating elements of this cyclic subgroup are a function of that particle's size\cite{fraleigh1982first}.

Using this mathematical model, which is valid for both the tetragonal and monoclinic lattices we previously described as well as all other periodic lattices, we can completely describe the trajectory of a particle advecting through an obstacle lattice after the third particle--obstacle interaction by determining the translational symmetry of the trajectory and if the trajectory is pure or mixed. For the former, we can describe the periodicity of the trajectories by a generator $\begin{bmatrix}a&b\end{bmatrix}^T$ representing the translational symmetry of the particle trajectories in lattice space, i.e. in terms of the location of the obstacle in lattice coordinates over which the trajectory repeats. Knowledge of the generator vector for a given particle/trajectory and how it transforms into real space by use of the transformation matrix $A$ enables quantitative statements about key particle advection-related quantities as detailed in Section \ref{transportquantities}.

\begin{figure*}
\centering
\includegraphics[width=0.9\textwidth]{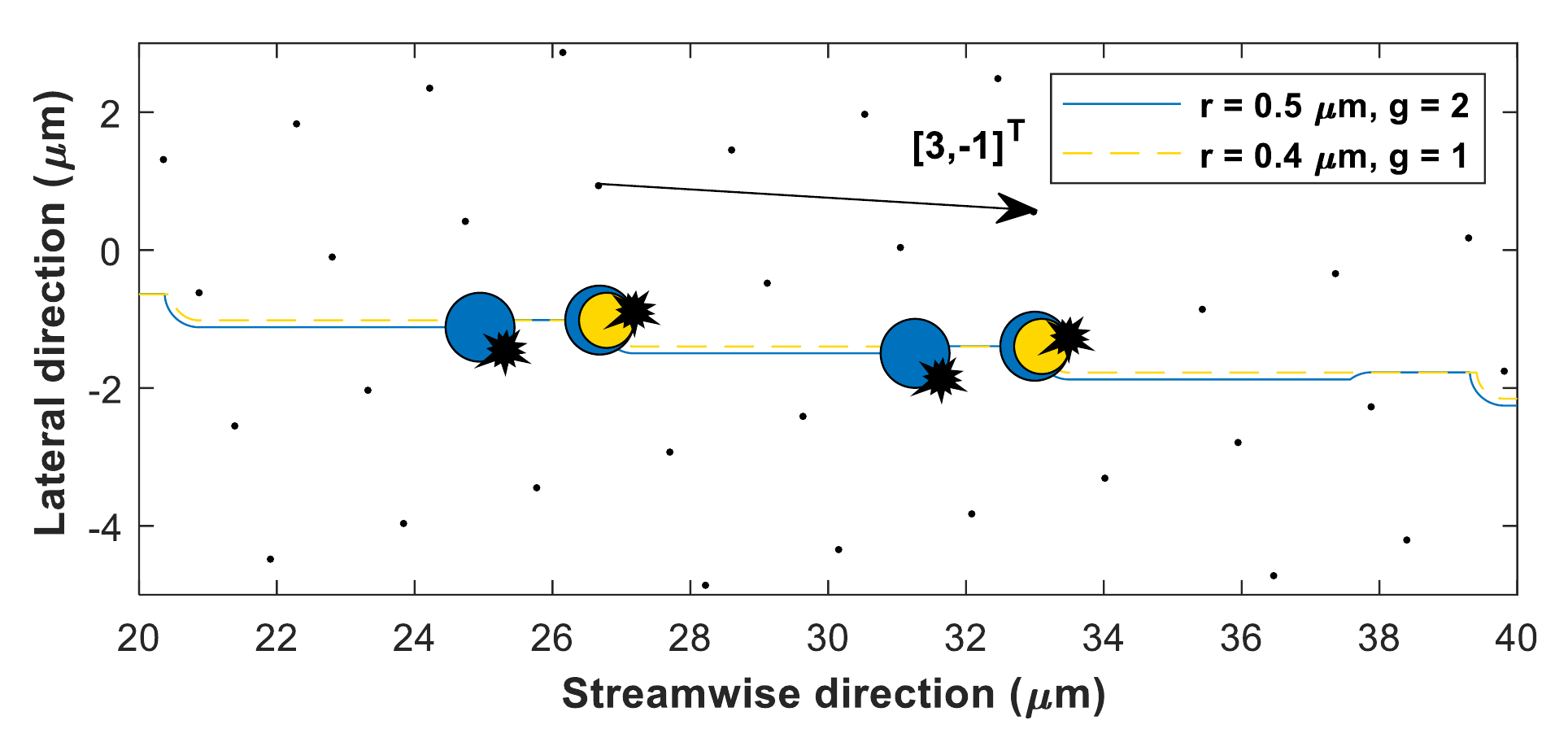}
\caption{\label{fig:internalcollisions} Trajectories of two differently-sized particles possessing the same generator through a square obstacle lattice, $\theta = 15^o$, $\Delta = 2\ \mu m$. Particles are drawn at example collision locations. The particle with the trajectory denoted in solid blue/dark grey ($r = 0.5\ \mu m$) collides twice per cycle with the obstacle array, whereas the particle with the trajectory denoted in dashed yellow/light grey ($r = 0.4\ \mu m$) collides once per cycle. As predicted by the arguments of Section \ref{traj}, these particles have the same lateral displacement per dynamical cycle but have twofold-different spatial collisions per cycle.}
\end{figure*}

For the latter, we restate that the translational symmetry associated with a specific particle trajectory does not uniquely specify how many particle--obstacle collisions occur within a single dynamical cycle. Thus as noted in \cite{PhysRevE.88.032136} \cite{PhysRevE.90.012302}, particles with a trajectory described by a generator $\begin{bmatrix}a&b\end{bmatrix}^T$ may have either one or two collisions per dynamical cycle, as implied by Figure \ref{fig:flowchart} using the pure/mixed notation and illustrated through simulation by Figure \ref{fig:internalcollisions}.

Trajectories for which one collision per cycle occurs have been termed ``fundamental'' \cite{PhysRevE.88.032136}, ``bump'' \cite{B515371A}, or ``displacement'' \cite{PhysRevE.78.046304} modes, whereas trajectories for which two collisions occur have been referred to as "zig-zag mode" \cite{B515371A} or "mixed mode" \cite{PhysRevE.88.032136} trajectories in the literature. The latter includes modes which have no net lateral displacement per cycle, and these zero-displacement modes have been the principal focus of mixed-trajectory analysis in the literature even though they form a small subset of this class of trajectories.

We now quantitatively analyze these trajectories by describing key advection-related quantities that produce the generator and the collisions per cycle for a particle of a given radius $r$ in an arbitrary lattice, and in turn generate functions describing lateral displacement and spatial collision frequencies as a function of particle size and lattice parameters.

\section{Key Transport Quantities\label{transportquantities}}

The symmetry-induced cyclical dynamics model shows that advecting particles will always quickly "lock in" to periodic trajectories, characterized by the lattice parameters governing the lattice periodicity, and that the periodic trajectory the particle eventually settles into is dependent on the particle's size owing to the repeated particle-obstacle interactions governed by it. We thus describe how we can use this mathematical model to extract key advection-related parameters such as the lateral displacement per length and spatial collision frequency of a particle as a function of the particle's size and of the lattice parameters.

\subsection{Critical Radius\label{critradius}}

Because the translational symmetry of a particle trajectory is directly related to the particle's size, one may anticipate the existence of "critical" radii beyond which particle trajectories will exhibit a different translational symmetry than those of a slightly larger particle. Consequently, we define a critical particle radius, which is the radius beyond which a particle cannot access a specific translational mode. Because we need consider only the location of the obstacle over which the trajectory repeats, we can calculate the critical radius for a given translational mode $\begin{bmatrix}a&b\end{bmatrix}^T$ by calculating the radius at which a particle that interacts with an obstacle in the origin fails to repeat its interaction with an obstacle in the lattice position  $\begin{bmatrix}a&b\end{bmatrix}^T$. For a downward traveling mode, this implies $-r_{\mathrm{crit}} = y$, and for an upwards traveling mode, $r_{\mathrm{crit}} = y$, where $y$ is defined from the lattice geometry using the equation in Section \ref{latticegeom}. We can combine both criteria into
\begin{equation}
r_{\text{crit}} = \left|y\right| = \left|a\left(\vec{l}_a\cdot\hat{y}\right) + b\left(\vec{l}_b\cdot\hat{y}\right)\right|
\end{equation}
The critical radius determines the location of the transitions between different translational modes, which are a prominent feature of the functions representing lateral displacement and spatial collision frequency shown below. It also provides information regarding the possibility of a particle advecting in a given translational mode, irrespective of whether it truly does so. Unlike the lateral displacement per length, a particle's spatial collision frequency can change discontinuously as a function of its radius for a fixed translational trajectory symmetry; we discuss this, and the radii at which those transitions occur as a function of the critical radii we derive here, in Section \ref{num_colls}.

\subsection{Collisions Per Cycle\label{num_colls}}

\begin{figure*}[h]
\centering
\includegraphics[width=0.9\textwidth]{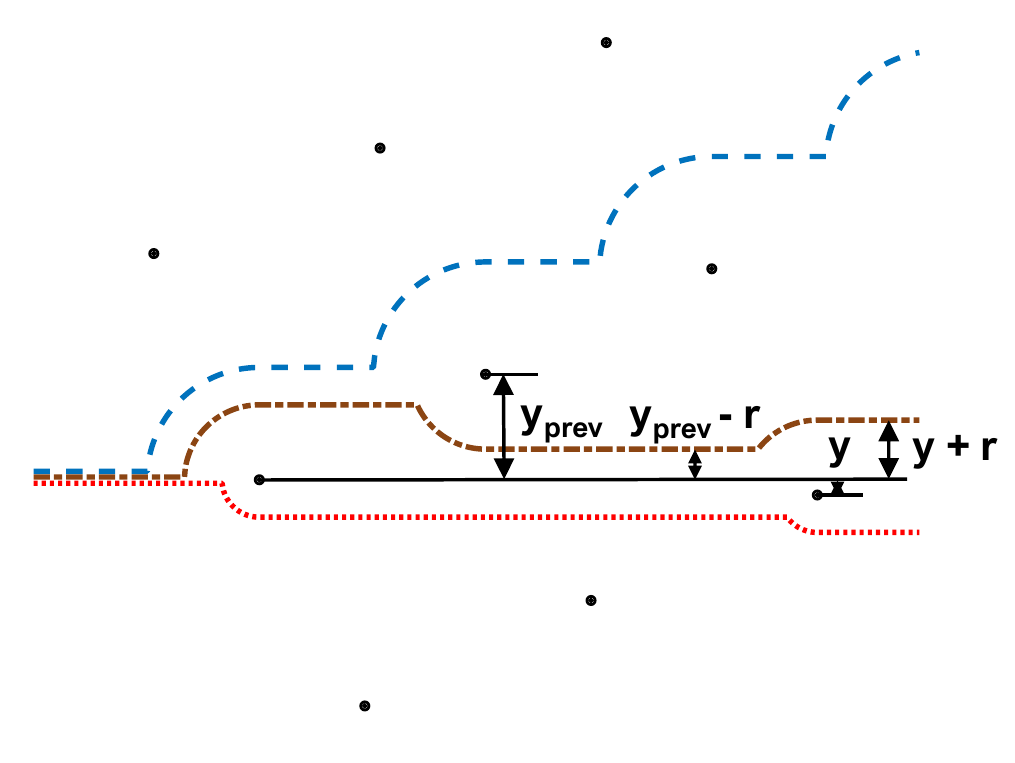}
\caption{\label{fig:mixedillustration}Two "pure" trajectories are shown with dashed/dotted lines, the trajectory associated with $y_\mathrm{prev}$ in dashed blue and the trajectory associated with $y$ in dotted red. For these two trajectories, a hypothetical "mixed" trajectory shown in dash-dotted brown results if \& only if $y_{\mathrm{prev}} - r \leq y + r$. Solid black lines have been added to guide the eye in measuring magnitudes of relevant quantities.}
\end{figure*}

In order to quantitatively understand the conditions under which a "pure" or "mixed" trajectory take place, we define a collision factor $g$ indicating the number of collisions a particle undergoes per dynamical cycle; 1 for a "pure" trajectory, and 2 for a "mixed" trajectory. By considering the geometry of the obstacle lattice and the formula for the critical radius in Section \ref{critradius}, we note that a test particle with a radius slightly smaller than the critical radius for a previously accessible mode $\begin{bmatrix}a_{\mathrm{prev}} & b_{\mathrm{prev}}\end{bmatrix}^T$ will still interact with the obstacle located at $\begin{bmatrix}x_{\mathrm{prev}} & y_{\mathrm{prev}}\end{bmatrix}^T$, but will be displaced laterally in a direction opposite to its previous direction. As a result, a particle of radius $r$ in a translational mode $\begin{bmatrix}a & b\end{bmatrix}^T$ interacts with this additional obstacle as long as $y_{\mathrm{prev}} - r \leq y + r$ for a net downward displacement mode or $y_{\mathrm{prev}} + r \geq y - r$ for a net upward displacement mode. See Figure \ref{fig:mixedillustration} for a diagram of the net downward displacement case.
Combining both inequalities, recalling the sequential sign difference between $y$'s, and using the definition of the critical radius stated above, we define the collision factor as:
\begin{equation}g = \begin{cases} 2 & r \geq \frac{|y-y_{\mathrm{prev}}|}{2} = \frac{r_{\mathrm{crit}_{\mathrm{prev}}} + r_{\mathrm{crit}}}{2} \\ 1 & r < \frac{|y-y_{\mathrm{prev}}|}{2} = \frac{r_{\mathrm{crit}_{\mathrm{prev}}} + r_{\mathrm{crit}}}{2}\end{cases}
\end{equation}
Once we possess information about all the possible translational modes as a function of particle size in a given lattice, and by extension the critical radii associated with these modes, we can directly calculate the collision factor as a function of particle size through our knowledge of the critical radii associated with those modes. For the mode with the smallest possible spatial frequency in the lattice (which is always a $\begin{bmatrix}1&0\end{bmatrix}^T$ or $\begin{bmatrix}0&1\end{bmatrix}^T$ mode by virtue of the lattice construction in Section \ref{latticegeom}), the collision factor $g$ is always $1$ because there is no "previous" obstacle for the particle to collide with. In addition, for a mode with purely streamwise translational symmetry, we find $g = 0$ when $r < \frac{r_{\mathrm{crit}_{\mathrm{\mathrm{prev}}}} + r_{\mathrm{crit}}}{2}$ because the particle will not interact with any obstacles after the first collision. Finally, it is possible that the $r_\mathrm{crit_\mathrm{prev}}$ associated with an $r_\mathrm{crit}$ is larger than the maximum radius $r_{\text{max}}$ allowed in the system. As long as these cases are accounted for, this collision factor $g$ describes the number of collisions per cycle in a given lattice completely given knowledge of the modes in the lattice as a function of size. This definition of $g$ leads directly into the definition of the spatial collision frequency below.

\subsection{Lateral Displacement per Unit Length\label{latdisp}}

If we approximate the trajectories of the particles as being straight lines between each obstacle over which the trajectories repeat, we can approximate the lateral displacement per unit length $\Upsilon$ of a particle in mode $\begin{bmatrix}a&b\end{bmatrix}^T$ as
\begin{equation}
\Upsilon = \frac{y}{x} = \frac{a\left(\vec{l}_a\cdot\hat{y}\right) + b\left(\vec{l}_b\cdot\hat{y}\right)}{a\left(\vec{l}_a\cdot\hat{x}\right) + b\left(\vec{l}_b\cdot\hat{x}\right)}
\end{equation}
where we can obtain $x$ and $y$ for a given mode $\begin{bmatrix}a&b\end{bmatrix}^T$ using the transformation matrix $A$ as discussed in Section \ref{latticegeom}. A related quantity is the effective transport angle $\alpha$, discussed in \cite{PhysRevLett.92.130602} \cite{risbud_drazer_2013}, which is simply $\alpha = \tan\frac{y}{x} = \tan\Upsilon$.

\subsection{Spatial Collision Frequency\label{collfreq}}

We can approximate the spatial collision frequency $\omega$ by obtaining the number of collisions a particle undergoes per dynamical cycle and dividing by the length of this trajectory:
\begin{equation}
\omega = \frac{g}{x} = \frac{g}{a\left(\vec{l}_a\cdot\hat{x}\right) + b\left(\vec{l}_b\cdot\hat{x}\right)}
\end{equation}
where $g$ is the collision factor discussed before and $x$ is again calculated for a given mode $\begin{bmatrix}a&b\end{bmatrix}^T$ using the transformation matrix $A$ as discussed in Section \ref{latticegeom}.

These parameters specify all relevant information about a trajectory given knowledge of its generator vector and the radius of the particle generating it. We conclude our analysis of this mathematical model by describing how to determine the generator associated with a given particle of radius $r$ advecting in a lattice with lattice vectors $\vec{l}_a$ \& $\vec{l}_b$.

\subsection{Constructing Lateral Displacement \& Collision Frequency Functions}

For a given particle radius and lattice geometry, it suffices to know only one generator parameter in order to obtain the other. This is because, for a given particle radius $r$ and suitably defined generator parameter $a$, we can straightforwardly identify the second generator parameter $b$ associated with the closest obstacle to an advecting particle in the $a$th lattice "row". To do so, consider a row of obstacles beginning from the origin in the direction of the lattice vector with the largest streamwise component, $\vec{l}_a$. As we increase the row number $a$, we increase the lateral distance of the obstacle from the origin $a\left(\vec{l}_a\cdot\hat{y}\right)$ until this lateral distance becomes larger in magnitude than that of the other lattice vector $\vec{l}_b$. As a result, once this criterion $\left|a\left(\vec{l}_a\cdot\hat{y}\right)\right| > \left|\left(\vec{l}_b\cdot\hat{y}\right)\right|$ is satisfied, the obstacle located at lattice position $(a,\pm 1)$ is closer to the origin laterally than the next obstacle in the original lattice row $(a,0)$. The spatial symmetry of the lattice then lets us continue applying this row-shifting rationale for any $a$, leading to an equation for the "column" position of an obstacle in the second lattice direction $b$ representing the closest obstacle to the origin laterally in a given row $a$:
\begin{equation}
b = \lfloor -a\left(\frac{\vec{l}_a\cdot\hat{y}}{\vec{l}_b\cdot\hat{y}}\right) \rceil
\end{equation}
where $\lfloor \rceil$ indicates rounding to the nearest integer. Because the mode the particle travels in has the smallest streamwise cycle length for those modes satisfying the collision criteria above, and because a particle can interact with at most two obstacles in a single "row", it suffices to sequentially test only two tentative generators $\left[a,-\lfloor b\left(\frac{\vec{l}_a\cdot\hat{y}}{\vec{l}_b\cdot\hat{y}}\right) \rceil - \text{sgn}\left(\vec{l}_a\cdot\hat{x}\right)\right]$,$\left[a,-\lfloor b\left(\frac{\vec{l}_a\cdot\hat{y}}{\vec{l}_b\cdot\hat{y}}\right) \rceil\right]$ associated with increasing lattice "row" number $a$ and assign to the particle the first generator that satisfies the collision criterion $\left|a \left(\vec{l}_a\cdot\hat{y}\right) + b\left(\vec{l}_b\cdot\hat{y}\right)\right| \leq r$. This process is far more computationally simpler, yet equivalent to, the optimization approaches for calculating translational modes in rotated squares proposed by \cite{PhysRevE.90.012302}, and for the mode selection criteria described in \cite{PhysRevLett.92.130602}, both of which use Euclidean distance rather than streamwise proximity.

\section{Lateral Displacement and Collision Frequency Functions\label{singleconstruct}}

Having described in the above sections a mathematical framework for calculating the lateral displacement and spatial collision frequency of a particle of given size advecting through a lattice with given lattice parameters, we now discuss how to construct functions of these transport quantities as a function of one of these parameters---particle size and lattice geometry---while fixing the other. These functions provide a quantitative metric for understanding the effects of altering lattice geometry on the advection behavior of the particles as well as insight into the size-dependence of these behaviors, and motivate the development of inverse design solutions for microdevices exploiting this behavior.

\subsection{Lateral Displacement and Collision Frequency as a Function of Array Parameters}
\begin{figure*}[h]
\centering
\includegraphics[width=0.9\textwidth]{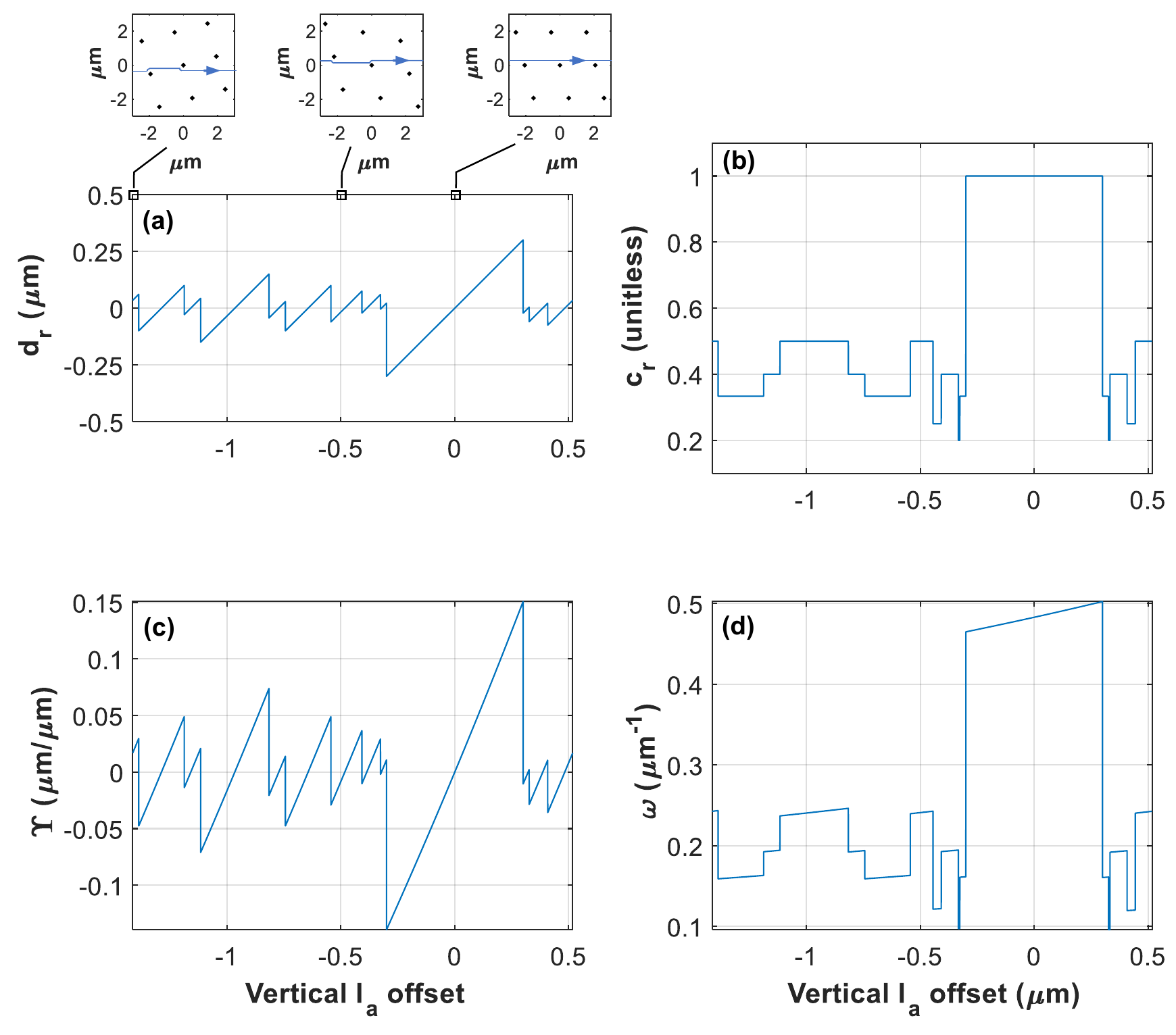}
\caption{\label{fig:aliasing} The top 2 subfigures represent lateral displacement per row $d_r$ (a) \& collision number per row $c_r$ (b) as a function of the vertical offset of the $\vec{l}_a$ lattice vector as a rotated square lattice is shifted along the $\vec{l}_b$ direction for a fixed particle radius of $0.3\ \mu\text{m}$. The lateral displacement per row functions are a linear combination of linear functions with unit slopes and offset $\frac{b}{a}(\vec{l}_b\cdot\hat{x})$. Collisions per row appear as linear combinations of constant functions with rational fraction values. Dividing by the appropriate lateral length yields the lateral displacement per length $\Upsilon$ (c) and spatial collision frequency $\omega$ (d). Diagrams of the lattice, as well as trajectory sketches of the $0.3\ \mu\text{m}$ particle through the lattice, are shown for specific offset values in (a).}
\end{figure*}

We can use the equations in Section \ref{transportquantities} to straightforwardly describe the way transport parameters are affected in general by changing any of the four components of the matrix $A$ describing the obstacle lattice. In particular, we consider the effect of shifting a lattice vector in the direction of another lattice vector, as this generates periodic functions of shift magnitude due to the intrinsic symmetry of lattices.
For a fixed particle radius, shifting either the $\vec{l}_a$ or $\vec{l}_b$ vectors in the direction of the other results in periodic lateral displacement per row functions of the shifted vector's lateral projection that are piecewise discontinuous linear functions, whose discontinuities arise as a result of modes becoming accessible/inaccessible in accordance with the criteria in Section \ref{critradius}. See the top left graph in Figure \ref{fig:aliasing} for an example of these. Each individual linear function has a unit slope and lateral intercept $\frac{b}{a}(\vec{l}_b\cdot\hat{y})$. 

Similarly, collisions per lattice row appear as piecewise discontinuous constant functions of magnitude $\frac{g}{a}$, where $a$ is the row number associated with the corresponding mode, for which an example is shown in the top right of Figure \ref{fig:aliasing}. The collision functions contain additional discontinuities in addition to those induced by mode transitions, coming from discontinuities in $g$ according to the criteria in Section \ref{num_colls}.

Multiplying these by the row number per streamwise length, $\frac{a}{a \left(\vec{l}_a\cdot\hat{x}\right) + b \left(\vec{l}_b\cdot\hat{x}\right)}$, results in lateral displacement per length and spatial collision frequency functions which are linear fractional functions; examples are shown on the bottom 2 graphs of Figure \ref{fig:aliasing}.

\subsubsection{Aliasing}
Although the graphs in Figure \ref{fig:aliasing} can be constructed entirely through use of the mathematical tools in Section \ref{transportquantities}, it is illustrative to show an alternative construction process that applies when ${\vec{l}_b\cdot\hat{y}}$ is sufficiently large that a particle can only interact with one obstacle within an obstacle row for fixed $a$.  Because the obstacle array simultaneously is (a) a periodic function, (b)  the actuator that controls the position of the particles (via displacement), and (c)  the sensor that detects the presence of particles (via observation of displacement or collisions), the complicated structure of the displacement and collision curves can be simplified by use of Fourier analysis and undersampling theory.  In this view, these graphs are generated by repeated aliasing of a fundamental spectral feature by an infinite set of shah functions corresponding to symmetries associated with all of the possible $a$ values proceeding to infinity. 


If we consider the plot of collisions per row $c_r$ vs. $\frac{\vec{l}_a\cdot\hat{y}}{\vec{l}_b\cdot\hat{y}}$ as a spectrum, the fundamental spectral feature corresponds to $\begin{bmatrix}a & b\end{bmatrix}^T=\begin{bmatrix}1 & 0\end{bmatrix}^T$; for an infinite array of infinitesimal obstacles, $c_r=1$ if $0 < \vec{l}_a \cdot \hat{y}< r$ and $c_r=0$ otherwise. Similarly, the lateral displacement of particles when advecting through the array can also be described for this fundamental mode, in that the lateral displacement per row $d_r$ normalized by $\vec{l}_b\cdot\hat{y}$ is given by $\frac{d_r}{\vec{l}_b\cdot\hat{y}}=\frac{\vec{l}_a\cdot\hat{y}}{\vec{l}_b\cdot\hat{y}}$ if $\vec{l}_a\cdot\hat{y}<r$ and zero if $\vec{l}_a\cdot\hat{y}>r$.


The threshold $\vec{l}_a\cdot\hat{y}>r$ corresponds to a transition away from $[a\ b]=[1\ 0]$; collision and displacement responses beyond this threshold can be predicted by considering the periodic nature of the rows as an infinite set of sampling functions and determining the Fourier aliasing that occurs when these sampling functions are applied to the fundamental mode.  A sampling function $\Sha(\vec{l}_a\cdot\hat{x})=\sum_{k=-\infty}^{\infty}\delta(x-k(\vec{l}_a\cdot\hat{x}))$, where $\delta$ is the Dirac delta function, can be used to describe the locations of the rows, where $x=a(\vec{l}_a\cdot\hat{x})$ and $a$ is an integer denoting the row number.


The distances $\vec{l}_a\cdot\hat{y}$, $\vec{l}_b\cdot\hat{y}$, and $\vec{l}_a\cdot\hat{x}$ relate to the periodic nature of the array (and the aliasing that results) in different ways. The spatial frequency $\Omega$ (in the $x$ direction) with which the obstacles are offset by $\vec{l}_b\cdot\hat{y}$ is $\Omega=\frac{2\pi(\vec{l}_a\cdot\hat{y})}{(\vec{l}_b\cdot\hat{y})(\vec{l}_a\cdot\hat{x})}$. Therefore, $\frac{\vec{l}_a\cdot\hat{y}}{\vec{l}_b\cdot\hat{y}}$ is proportional to the spatial \textit{frequency} of the shifting.  The shifting is then sampled by each row at a spatial frequency $\Omega_S$ given by $\Omega_S=\frac{2\pi}{\vec{l}_a\cdot\hat{x}}$;  $\vec{l}_a\cdot\hat{x}$ is thus the spatial \textit{period} of the sampling.  For a signal with frequency $\Omega$ sampled at frequency $\Omega_S$, the result is consistent with a signal at any frequency $\Omega'=\Omega-N\Omega_S$, where $N$ is an integer.  This leads to an aliasing result in which an offset $\vec{l}_a\cdot\hat{y}$ is equivalent to an offset given by $\vec{l}_a\cdot\hat{y}-N\vec{l}_b\cdot\hat{y}$ or, because of the symmetry of the Fourier transform around zero for real functions, the result for an offset $\vec{l}_a\cdot\hat{y}$ is also equivalent to an offset given by $N\vec{l}_b\cdot\hat{y}-\vec{l}_a\cdot\hat{y}$.  Because of this property, we need only look at the collision and displacement responses between $0<\vec{l}_a\cdot\hat{y}<\vec{l}_b\cdot\hat{y}$. The responses are symmetric (for collisions) or antisymmetric (for displacement) when reflected over the Nyquist or folding frequency $\vec{l}_a\cdot\hat{y}_N=\vec{l}_b\cdot\hat{y}/2$.

The $c_r$ and $d_r$ curves also contain contributions from particles that do not hit every row; this corresponds to sampling at $\frac{1}{2}\Omega_S$, $\frac{1}{3}\Omega_S$, ... $\frac{1}{p}\Omega_S$.  For collisions, the response of modes that hit every $p$ rows have peak $c_r$ values given by $\frac{g}{p}$.  These samples alias as well, so that any signal at frequency $\Omega$ is aliased at magnitude $\frac{g}{p}$ at all spatial frequencies
$\Omega-\frac{N\Omega_S}{p}$,
for all integer $N$ and $p$.  For an infinitely long array, the aliased modes for $p=1,2,\ldots,\infty$ have amplitude $g/p$ in the range $\frac{q\vec{l}_b\cdot\hat{y}}{p}<\frac{\vec{l}_a\cdot\hat{y}}{\vec{l}_b\cdot\hat{y}}<\frac{a+q\vec{l}_b\cdot\hat{y}}{p}$, for all $q=0,1,\ldots,a-1$.  Thus, the aliased modes have peak magnitude $g/p$ and a width given by $r/p$. The observed $c_r$ curve exhibits the \textit{maximum} of these modes, because a particle that collides after a small number of rows never has an opportunity to collide (\textit{i.e.}, be sampled by) later rows of obstacles.

The displacement response can be constructed with the same aliasing approach.  The aliased modes have the same slope as the fundamental mode and width given by $r/a$.  As a result, the displacement response is simpler than the collisional response---the displacement response for infinite arrays is piecewise linear with unit slope at all points except for discontinuities.  These discontinuities correspond to changes in the number of rows in the mode, but not the exact nature of the collisions, so collision factor transitions do not generate discontinuities in the displacement graph (though they naturally generate discontinuities in the collision rate).  The displacement graph also exhibits geometric structure---the lines $d_r=-\frac{\vec{l}_a\cdot\hat{y}}{\vec{l}_b\cdot\hat{y}}\frac{r^2}{\left(\vec{l}_b\cdot\hat{y}-r\right)^2}$
and $d_r=\left(1-\frac{\vec{l}_a\cdot\hat{y}}{\vec{l}_b\cdot\hat{y}}\right)\frac{r^2}{\left(\vec{l}_b\cdot\hat{y}-r\right)^2}$ intersect the termini of the linear components of the response.

\subsection{Lateral Displacement and Collision Frequency as a Function of Particle Size}

Here we focus on constructing lateral displacement and spatial collision frequency functions of particle size for a given lattice geometry. In this scenario, the generators associated with each particle radius are fixed because the lattice geometry is fixed; hence the problem simplifies into finding the generators and collision factors associated with a range of particle radii, and calculating the transport quantities associated with them with the expressions from Section \ref{transportquantities}. 

The process of determining lateral displacement per length and spatial collision frequency as a function of particle size is considerably simpler than the equivalent process for functions of lattice parameters. This is as a result of the fact that the mode transitions determined in Section \ref{critradius} and collision transitions determined in Section \ref{num_colls} are directly based on radial criteria. As a result, functions of lateral displacement per length $\Upsilon(r)$ and spatial collision frequency $\omega(r)$ will appear as piecewise discontinuous constant functions that transition at the $r_{\text{crit}}$ (and $\frac{r_{\mathrm{crit}_{\mathrm{prev}}} + r_{\mathrm{crit}}}{2}$ for collisions) associated with that translational mode (see Figure \ref{fig:combinedsquares}). In addition, lattices that do not possess translational symmetry in the streamwise direction will contain an infinite amount of nontrivial generators---and by extension an infinite number of accessible transport modes---leading to self-similar structure in these functions as $r \rightarrow 0$.

As the traversed modes are those modes which are both accessible ($r_{\mathrm{crit}} > |y|$) and which possess the smallest streamwise cycle length, the lateral displacement magnitude per unit length $(|\Upsilon|)$ as a function of particle size monotonically increases as $r$ increases. This is not the case for the collision frequency in general, because of the discontinuities present in $g$ that are distinct from the generator discontinuities.

\subsection{Transport Functions in Specific Lattices}
\subsubsection{Rotated Squares\label{rotsqeq}}

\begin{figure*}[h]
\centering
\includegraphics[width=0.9\textwidth]{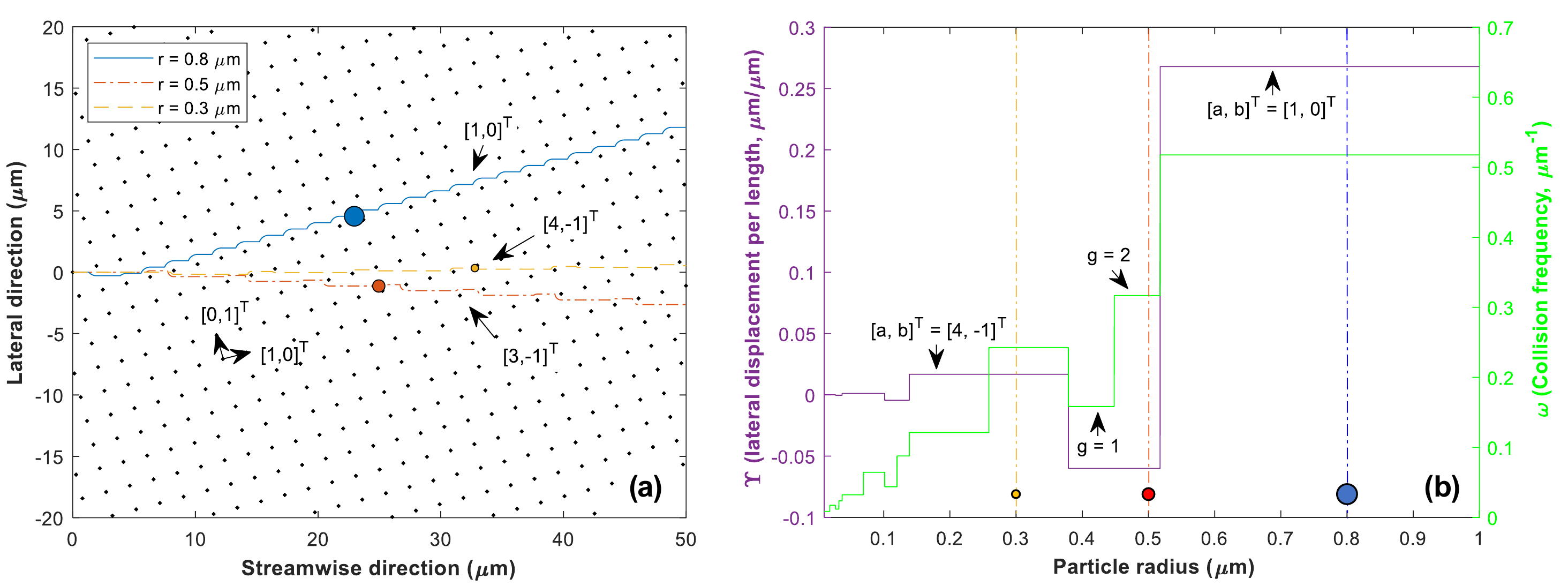}
\caption{\label{fig:combinedsquares} Trajectories of differently-sized particles through a rotated square obstacle array and their corresponding generators for $r = 0.8\ \mu\text{m}$ (solid blue), $r = 0.5\ \mu\text{m}$ (dash-dot red), and $r = 0.3\ \mu\text{m}$ (dashed yellow) are shown in (a). Lateral displacement per length ($\Upsilon$, purple/dark grey) and collision frequency ($\omega$, green/light grey) functions of particle size for this lattice as predicted by the arguments of Section \ref{latdisp} and \ref{num_colls} are shown in (b); dash-dotted lines in (b) correspond to the particle sizes whose trajectories are represented on the left panel. Note the presence of differing collision frequencies for a given lateral displacement per length in all modes except for the primary (right-most) mode. For this lattice, $\theta = 15^o$ and $\Delta = 2\ \mu\text{m}$.}
\end{figure*}

For a square lattice, we apply the geometric definitions for the lattice described in Section \ref{latticegeom} to the equations shown in \ref{transportquantities} to obtain the following specific expressions for transport quantities in rotated square lattices. Firstly, the critical radius for a particle in some mode $\begin{bmatrix}a&b\end{bmatrix}^T$ advecting through a square obstacle lattice is:
\begin{equation}
r_{\mathrm{crit}_{\mathrm{s.l.}}} = \Delta|a\sin(\theta) +b\cos(\theta)|
\end{equation}
For the lateral displacement per unit length, we find:
\begin{equation}
\Upsilon_{\mathrm{s.l.}} = \frac{a\sin(\theta) + b\cos(\theta)}{a\cos(\theta) - b\sin(\theta)}
\end{equation}
For the collision frequency, we find:
\begin{equation}
\omega_{\mathrm{s.l.}} = \frac{g}{\Delta(a\cos(\theta) - b\sin(\theta))}
\end{equation}
The predictions for $r_{\mathrm{crit}_{\mathrm{s.l.}}}$ and $\Upsilon_{\mathrm{s.l.}}$ are consistent with the directional locking model described in \cite{PhysRevE.90.012302} in the limit where the obstacle size becomes infinitesimally small and with the deterministic model described in \cite{PhysRevLett.92.130602}. Plots of these last two quantities as functions of particle size along with sample particle trajectories are shown in Fig. \ref{fig:combinedsquares} for an example lattice of this type.

\subsubsection{Oblique Lattices}

For a lattice composed of oblique lattices, we proceed as above by applying the geometric definitions for the lattice described in Section \ref{latticegeom} to the equations shown in \ref{transportquantities} to find the specific expressions for transport quantities in oblique lattices.

The critical radius for a particle in some mode $\begin{bmatrix}a&b\end{bmatrix}^T$ advecting through an oblique obstacle lattice is:
\begin{equation}
r_{\mathrm{crit}_{\mathrm{o.l.}}} = |a\Pi - b\Gamma|
\end{equation}
For the lateral displacement per unit length, we find:
\begin{equation}
\Upsilon_{\mathrm{o.l.}} = \frac{a\Pi - b\Gamma}{a\Lambda} = \frac{1}{\Lambda}\left({\Pi - \frac{b\Gamma}{a}}\right)
\end{equation}
For the collision frequency, we find:
\begin{equation}
\omega_{\mathrm{o.l.}} = \frac{g}{a\Lambda}
\end{equation}
The prediction of $\Upsilon_{\mathrm{o.l.}}$ for the first mode is equivalent to that described by the recurrence-map model for the particle migration angle in \cite{Kim201706645} for point-like obstacles, the prediction of $\omega_{\mathrm{o.l.}}$ for the first mode is equivalent to the ballistic collision model in \cite{PhysRevE.88.032136}, and the predictions for both of these are in agreement with advection simulations in \cite{PhysRevE.88.032136}.

\section{Chained Obstacle Lattices and Inverse Design Schemes}

Having obtained a full quantitative model for the advection of particles through an obstacle lattice, we now consider the behavior of these particles through a sequential chain of obstacle lattices with differing lattice parameters. These chained lattices were initially considered in \cite{Huang987} and are used for both displacement-based and collision-based\cite{10.1371/journal.pone.0035976} applications. These chained obstacle lattices generate advection behaviors that cannot be achieved through individual lattices, and we show below that such devices can be designed to approximate any physically reasonable lateral displacement function of particle size to arbitrary accuracy. 

By neglecting entrance effects (which include particle motion before the translational mode lock-in), we can approximate the total lateral displacement and collision number through a chain of lattices to be the sum of these quantities through each of the individual lattices. Denoting the total lateral displacement through the chain as $d$ and the total collision number as $c$, we can express the statement above mathematically as: 
\begin{subequations}
\begin{align}
d = \sum_{i=1}^{N} d_{i} = \sum_{i=1}^{N} \ell_{i}\Upsilon_{i} \\
c = \sum_{i=1}^{N} c_{i} = \sum_{i=1}^{N} \ell_{i}\omega_{i}
\end{align}
\end{subequations}
where $d_{i}$ and $c_{i}$ are the lateral displacement and collision number through each individual lattice, and $\ell_i$ represents the streamwise length of each individual lattice in the chain. With this approximation in mind, we can straightforwardly construct the transport quantity functions of either size or lattice parameters for the lattice chains using the equations above and the formulas we obtained in Section \ref{singleconstruct}.

\subsection{Approximating Displacement \& Collision Functions of Particle Size}

Having described a mathematical framework for understanding the collision frequencies and lateral displacements per length of particles advecting through chained obstacle lattices, we now discuss how we can use "chained" obstacle lattices to solve the inverse design problem of approximating a desired size-dependent lateral displacement or collision frequency function. For the purposes of the following analyses, we will focus specifically on the rotated square lattices described in Section \ref{rotsq} and Section \ref{rotsqeq}, as this family of lattices exhibits isotropic fluid permeability (see \cite{C7LC00785J}) which minimizes performance degradation due to induced lateral pressure gradients. However, our analyses are straightforwardly extendable to any set of obstacle lattices where either $\vec{l}_a\cdot\hat{y}$ or $\vec{l}_b\cdot\hat{y}$ can be varied continuously from $0$ to the largest radius $r_{\max}$ of a particle flowing through the lattice without violating the no-clogging constraint.

For the inverse design problem in which we seek to approximate a given lateral displacement function of particle size, we demonstrate that the problem is solvable within an arbitrary degree of accuracy by showing that we can use displacement functions of size in chained obstacle lattices---which are linear combinations of the functions representing lateral displacement in individual rotated square obstacle lattices---to approximate any "reasonable" size-dependent lateral displacement function to arbitrary accuracy using a variety of standard optimization metrics. More formally, we prove that the underlying set of the multiset of functions representing lateral displacement through these lattices, which we will denote as $\mathbb{D}$, is dense on the space of $L^{1\leq p\leq \infty}$ functions defined on the closed interval of all possible particle radii up to some arbitrarily large radius $r_{\max}$, denoted as $L^{1\leq p\leq \infty} [0, r_{\max}]$.

For the inverse design problem in which we seek to approximate a given collision number function of particle size, we find that constructing optimization routines with guaranteed convergence is nontrivial due to the strictly non-negative nature of collision frequency/number. This can be intuited from the additive nature of the collision frequency functions---any overshoot in the collision number function can only become larger because there is no way to decrease the total collision number using additional chained arrays. As a result, the rest of this paper is focused on the inverse design problem associated with lateral displacement.

Because rotated square lattice angles and spacings can be chosen to make the lattice symmetric in the streamwise direction, the displacement of a particle can be made to have a specific single-size threshold, making the lateral displacement functions Heaviside functions which can then be superposed to make any lateral displacement function.  Formally, at specific rotation angles $\theta^*_{n} = \tan^{-1}(\frac{1}{n})$ where $n \in \mathbb{Z}\setminus \{-1,1\}$, the lattice obtains discrete translational symmetry in the streamwise direction and the only accessible modes are one mode with non-zero lateral displacement, when $r>r_{c}$, and a mode with no lateral displacement due to the alignment of the translational symmetry vector with the streamwise axis. Consequently, the lateral displacement per unit length functions become multiples of Heaviside step functions of the form $\frac{1}{n} H\Big(r-\frac{\Delta}{\sqrt{(n^{2}+1)}}\Big)$. Altering the obstacle spacing for these functions is thus equivalent to horizontal shifting of the function, and  we can therefore translate the non-zero part of the function anywhere in the interval $\Big(\frac{2r_{\max}}{\sqrt{(n^{2}+1)}},r_{\max}\Big]$ where the lower bound comes from the no-clogging constraint on the obstacle spacing. By using the simple function approximation theorem\cite{stein2009real}, the closed linear span of the set of functions with $\theta = \theta^*_{n}$ and $\Delta>2r_{\max}$ thus includes $L^{1\leq p\leq \infty}\Big(\frac{2r_{\max}}{\sqrt{(n^{2}+1)}},r_{\max}\Big]$. By increasing $n$ to an arbitrarily large integer, we can make the lower bound (and magnitude of the nonzero part) arbitrarily close to zero, and thus the closed linear span of our family of functions includes $L^{1\leq p\leq \infty}(0,\infty]$ as $n \rightarrow \infty$. Finally, we add to the function family a constant lateral displacement function over the entire interval so that the closed linear span includes $L^{1\leq p\leq \infty}[0,r_{\max}]$, concluding our argument. Such a constant lateral displacement function physically represents a lateral shift of the microdevice outlet relative to the inlet.

This result indicates that we can use a chain of rotated square obstacle lattices to approximate any size-dependent lateral displacement function to arbitrarily small error, where the error can be defined using many well-known optimization metrics. For example, this proof shows that we can devise an algorithm that finds a series of obstacle lattices approximating any target function by minimizing the sum of absolute errors between the approximation and the target ($p=1$), the sum of squared errors ($p=2$), or by minimizing the maximum absolute error ($p=\infty$). However, this proof does not outline any specific method by which to generate these approximations/device designs, nor does it indicate which method of these is best. In particular, the method present by the above proof is typically far from optimal.

\section{Approximation Schemes\label{approxschemes}}

Although the previous section shows that chained obstacle lattices can approximate any lateral displacement function of particle size in $L^{1\leq p\leq \infty}[0,r_{\max}]$ to arbitrary accuracy, it does not detail design procedures through which this can be done. Several techniques that can be used for this purpose are listed below.

\subsection{Na{\"i}ve Riemann Approximation}

An intuitive design approach involves a na{\"i}ve Riemann approximation, which resembles the approximation of an integral by use of rectangle functions.  This approach uses only a small subset of the possible object functions, but is amenable to intuitive design.

Informally, the na{\"i}ve Riemann approximation technique uses the fact that there are a set of lattice parameters that generate step functions in the size--displacement response, and adds these step functions by chaining lattices serially in the device to obtain an approximation of the desired target function.  Formally, using the lattice framework described in this paper, the na{\"i}ve Riemann approximation technique with $m$ lattices consists of the following steps:
\begin{enumerate}
\item Determine the largest rotation angle $\theta^*$ such that (a) the corresponding lateral displacement function is a step function ($\theta^*_{n} = \tan^{-1}(\frac{1}{n})$) and (b)  the location of its mode transition can be moved anywhere on the spectrum of particle sizes in the suspension of interest without interfering with the no-clogging constraint. Each individual lattice will possess this rotation angle.
\item Identify $m$ target points/radii $r^*$ of interest on the spectrum of particle sizes in the suspension, where the lateral displacement function will have a discontinuity.
\item For each of the $m$ target points $r^*$, obtain a  spacing values $\Delta^*$ such that the mode transition occurs precisely at $r^*$ for spacing $\Delta^*$ and angle $\theta^*$.
\item Alter the lengths $\ell^*$ (and, if needed, signs of $\theta^*$) of each lattice such that the value of the total lateral displacement function $d$ immediately after the discontinuity matches the target function at $r^*$.
\end{enumerate}

The na{\"i}ve Riemann approach requires no optimization algorithms and will eventually converge onto the target function as the number of points $r^*$ increases.  Further, this algorithm uses one lattice for each size threshold, and thus a set of $n$ particle populations can be separated through use of $n-1$ straightforwardly designed lattices. However, this approach has its drawbacks; as a result of fitting the function only to a discrete set of particle sizes $r^*$, it will intrinsically generate designs oriented exclusively around specific particle radii and not to a spectrum of particle sizes. Moreover, this approach is inherently inefficient for continuously polydisperse suspensions, because the lattice rotation angles are small and each lattice contributes only to thresholding in one local size range.  Consequently, these designs can be expected to under-perform when attempting to manipulate polydisperse suspensions. In addition, the presence of sharp features in the target function can lead to considerable aliasing unless the target points are chosen carefully, preventing the development of an efficient automated design algorithm exploiting this procedure.

\subsection{$L^2$ Optimization\label{l2opti}}
Although the na{\"i}ve Riemann approach is straightforward, as described above it underperforms for most functions.  By using all available lattice functions, most target functions can be approximated with shorter device length, fewer lattices, and more accuracy.  The set of non-monotonic lattice functions, however, require a more subtle design approach because each function may have more than one discontinuity and more than one amplitude.

Just as a bounded function can be described with a Fourier series by the sum of a set of sinusoids with magnitudes defined by the inner product of the target function with those sinusoids, the target function can be described by a sum of a set of lattice lateral displacement functions whose magnitudes are defined by the inner product of the target function with the lateral displacement function. This problem is more complicated because (a) the lateral displacement functions form a multiset, so there are duplicate functions and (b) the engineering design for these devices is an optimization of system error, device length, and design simplicity rather than simply error.

To this end, we mimic the process of Fourier series approximation by using an $L^2$ optimization process to sequentially
\begin{enumerate}
\item Identify the individual lateral displacement function that best approximates the target function.
\item Append the lattice corresponding to the function found above to the device design.
\item Replace the target function with the previous target function minus the newly found lateral displacement function.
\end{enumerate}

More formally, we could sequentially find the lateral displacement function of an individual lattice $d_i(\theta,\Delta,\ell)$ that best fits the target function $T$  in the least-squares sense by minimizing the $L^2$ norm of the vector rejection $\|T - d_i(\theta,\Delta,\ell)\|$ (hereafter referred to as the error). We would then subtract it from the target function, and continue procedurally using a new target function $T - d_i$ until the error in the approximation $d = \sum_{i=1}^{n} d_{i}$ is sufficiently small. This approach is particularly attractive---because $L^2$ is a Hilbert space, this error is guaranteed to decrease monotonically due to the same subspace exclusion arguments used to derive the Hilbert projection theorem\cite{stein2009real}. 

\begin{figure}[h]
\centering
\includegraphics[width=0.9\textwidth]{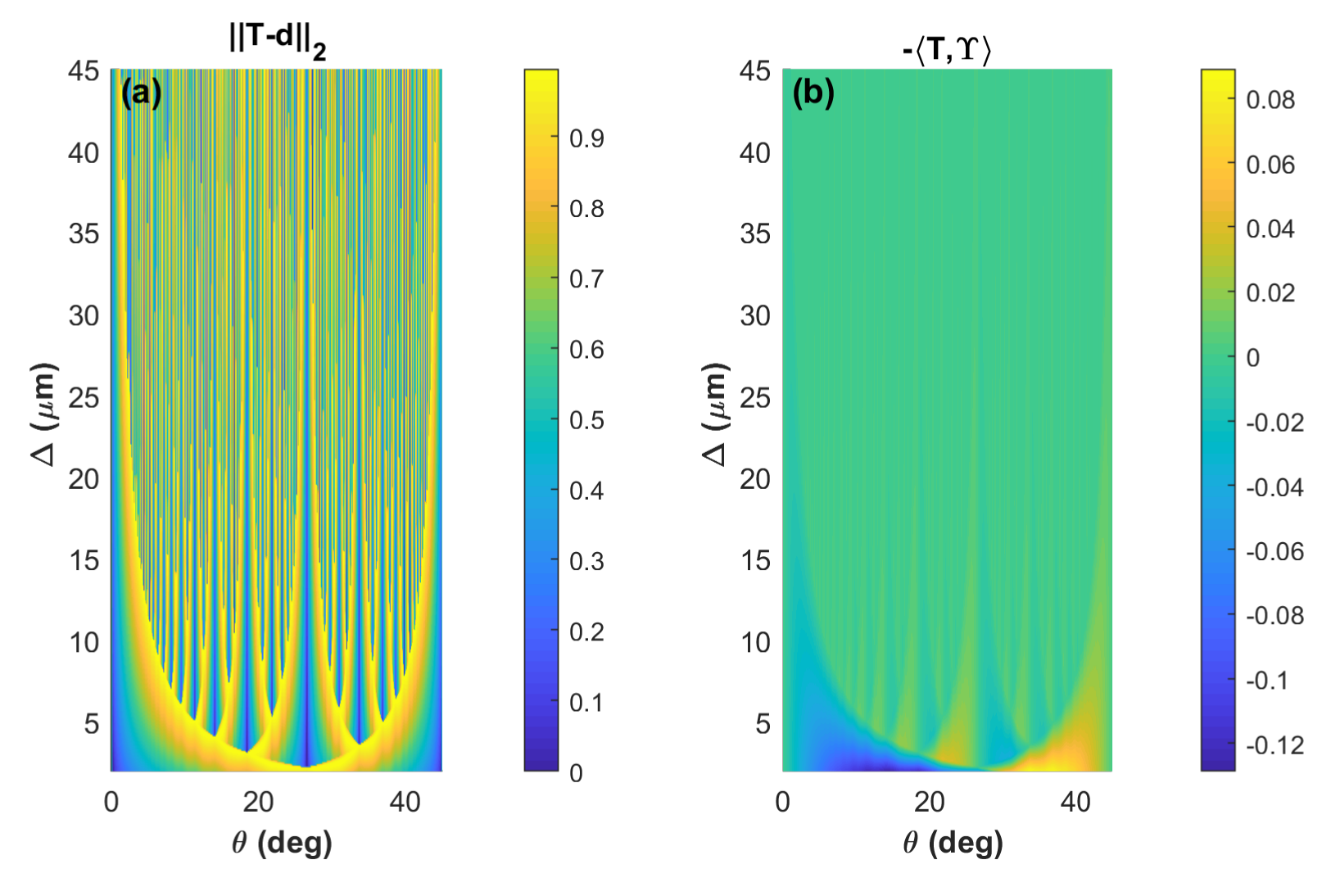}
\caption{\label{fig:residual} Comparison between the objective functions of $L_2$ error optimization (a) and the inner product method (b) in the case where $T$ is a constant unit-valued function over the particle size interval $[0.1, 1]$. Minima (in dark blue/grey) correspond to optimal design parameters. Note the presence of global minima inside the self-similar structures of the $L_2$ error objective function; these global minima are not present in the inner product method function, and are replaced by a single global minimum in a region devoid of self-similar structures (which are now maxima). The length associated with each $[\theta, \Delta]$ pair in the $L_2$ error function is calculated using the same procedure detailed in Section \ref{l2opti}.}
\end{figure}

The specific lateral displacement functions we choose are important.  Because this collection of functions comprises a multiset which contains an infinite number of duplicates, a cost function based only minimizing the error can (and often does) generate an infinite number of global and/or local minima (see Fig. \ref{fig:residual}), violating the principal uniqueness result of the Hilbert projection theorem. In addition, this approach minimizes error with respect to the target function, but does not address engineering concerns, specifically the minimization of design complexity and minimization of device length.  Thus our method must choose the lateral displacement functions in a manner that minimizes device length and complexity.

We achieve these engineering optimizations by maximizing the inner product between the lateral displacement per length functions and the target function of interest. Specifically, we evaluate the rate of change of error versus length of a lateral displacement function in the limit when $\ell$ goes to zero, $\frac{\partial\|T-d_i(\theta,\Delta)\|}{\partial\ell}_{\ell\rightarrow0}$, as our quantity of interest to minimize. Effectively, this minimization identifies the lattice lateral displacement function that effects the desired displacement in the shortest device length. By differentiating under the integral sign, minimizing this quantity is equivalent to minimizing $-\langle T,\Upsilon(\theta,\Delta) \rangle$, the negative of the inner product of the target function and the candidate lateral displacement per length functions. This straightforwardly removes the existence of the main source of multiple global critical points; solutions with different device lengths. (See Fig. \ref{fig:residual} for a comparison using direct evaluations of each cost functions discussed above.) 

Maximizing the inner product does not, however, remove the existence of the potentially infinite local minima/maxima; this combined with the highly nonlinear dependence of $\Upsilon$ on $\theta$ and $\Delta$ indicate a metaheuristic optimization algorithm is best suited for solving this optimization problem.

Equipped with this cost function, we can exploit this mathematical structure of the functions $d_i$ to calculate the values of each individual lattice length that best matches the target function for a given set of $d_{i}(\theta,\Delta)$ through the following method:
\begin{enumerate}
\item Find the $[\theta,\Delta]$ pair such that its corresponding $\Upsilon(r)$ function minimizes our cost function using a genetic optimization algorithm\cite{Goldberg:1989:GAS:534133}.
\item Calculate the best $\ell_{i}$ w.r.t. minimizing the $L^{2}$ norm of the error for the selected $\Upsilon(r)$ functions. We can calculate this without using any optimization scheme by exploiting that, because of the linear dependence of $d$ on $\ell$, this is a linear least-squares problem with closed-form solution $\ell_{i} = D^{+}T$, where $D^{+}$ is the Moore--Penrose pseudoinverse operator of the quasimatrix\cite{8178963} with columns $\Upsilon_{i}(r)$.
\item Continue the process iteratively using the new target curve $T_\mathrm{new} = T - d(r)$.
\end{enumerate}
As mentioned previously, because arguments similar to those in the projection theorem ensure that every $T_\mathrm{new}$ curve is orthogonal to $d(r)$, we ensure the error in successive approximations can only decrease as individual lattices are added. In addition, our choice of cost function and the linear dependence of lateral displacement functions $d$ on $\ell$ means we only have to conduct an optimization search over $[\theta,\Delta]$, reducing the dimensionality of the optimization problem from $\frac{3}{2}(n^2 + n)$ to $2n$ dimensions for a design with $n$ individual lattices.
\subsubsection{$\theta$-Restricted $L^2$ Optimization}
The approach above can be modified by fixing the angle $\theta$ of all the lattices in the chain to one corresponding to a fundamental mode using the same criteria detailed in Step 1 of Section 5.1. This approach is also guaranteed to have a monotonic error decrease, with the added benefit that one only needs to conduct an optimization search over $\Delta$ as the angle is already selected for each lattice, improving algorithm solving times considerably. However, it possesses the drawback (along with the na{\"i}ve Riemann approach) that device lengths automatically become quite large for devices expected to handle distributions with large particle size differences.

\section{Approximation Scheme Comparisons and Design Efficiency Measures}
These design approaches generate system designs, each of which can be evaluated by a number of metrics.  We choose three particular design metrics.  First, the mean square error $e$, indicates the mean square error ($L^2$ residual) between the target function and the device lateral displacement function.  This evaluates how well the design replicates the desired target function.  In addition to error, we consider both the physical device length and what we term the device complexity.  The total physical device length $\ell_{\text{tot}} = \sum_{i=1}^N \ell_i$, is calculated as the sum of each streamwise lattice length in a device design. Long devices exhibit diffusive effects and make the device sensitive to uncertainties in input flow rates. We describe the device complexity $N$, as the total number of distinct lattices in a given microdevice design.  This is ideally to be minimized, as more lattices requires more CAD design time and introduce more opportunities for edge effects.

The minimization of each of these metrics represents a reduction of distinct sources of error in an experimental actualization of such a device. Minimizing the total device length $\ell_{\text{tot}}$, for example, reduces the effects of diffusive lateral translation, whereas minimizing the device complexity $N$ reduces entrance effects associated with lateral particle translations from aperiodic particle-obstacle interactions, i.e. before the particle is "mode-locked". Using the information from Section \ref{approxschemes} and the definitions for these design metrics, we can compare some key characteristics of our approximation schemes with respect to these design metrics, shown in Table \ref{tab:approxcharacteristics}.

In total, these three metrics formulate the design task as follows: approximate the target function as well as possible in the shortest possible device with the fewest number of lattices.  We illustrate the outputs of these algorithms by displaying lateral displacement functions relative to the target function, and we evaluate different design algorithms by plotting mean square error versus device length---optimal performance corresponds to the lower-left portion of the graph.  We display the effects of complexity by showing each algorithm as a path through this error--length space as the number of lattices is increased.

\begin{table*}
\centering
\begin{ruledtabular}
\begin{tabular}{lcc}
\hline
& Na{\"i}ve Riemann & Optimization \\\hline
Error vs. Complexity & Can Increase/Decrease & Always Decreases \\
Error vs. Length & Can Increase/Decrease & Always Decreases \\
Length vs. Complexity & Can Increase/Decrease & Can Increase/Decrease \\
Optimization Procedure & Not Required & Required \\
Maximum Complexity & No. of Data Points & No. of Data Points - 1 \\
\end{tabular}
\end{ruledtabular}
\caption{\label{tab:approxcharacteristics} Tabular comparison of key characteristics of the na{\"i}ve Riemann approximation method and both optimization (direct $L^2$ and $\theta$-restricted $L^2$) methods.}
\end{table*}

\begin{figure*}[h]
\centering
\includegraphics[width=0.95\textwidth]{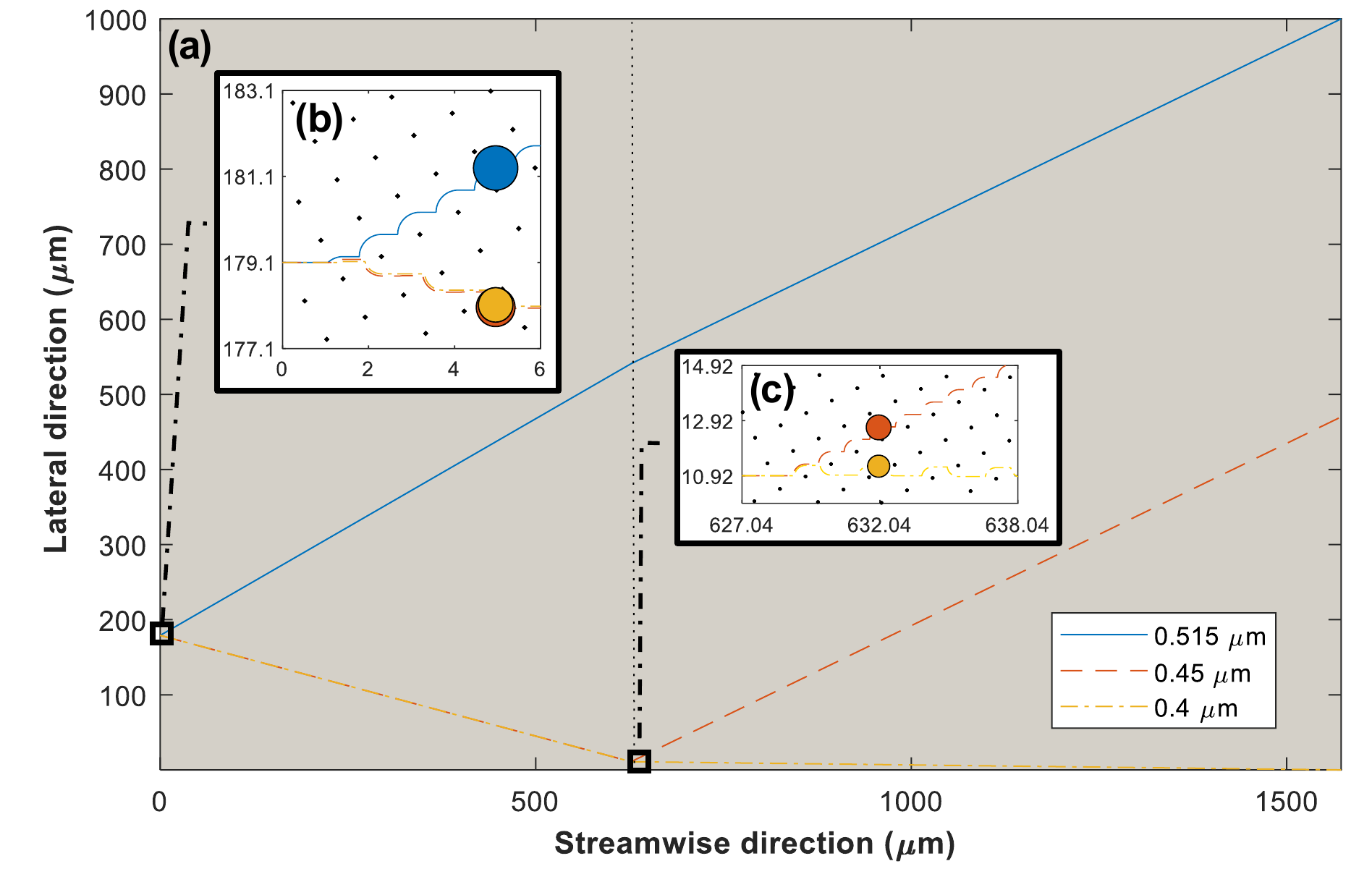}
\caption{\label{fig:huangdevice2} A size-based particle sorting microdevice (a), consisting of two chained obstacle arrays separated by a dotted line, designed using the direct $L_2$ technique to match the relative lateral displacements of particles reported in \cite{Huang987}. The trajectories of particles through the device for the approximate particle sizes used in \cite{Huang987} are also shown in (a). The light gray background represents the obstacle arrays, which are shown as points in the inset graphs (b) and (c). The device exploits multiple modes with large lateral displacements per lengths simultaneously to achieve sorting efficiency. The location of the inlet where particles are inserted is also determined by the $L_2$ algorithm. A comparison between this device and the device in \cite{Huang987}, is shown in Table \ref{tab:huangcomparison}. The lattice parameters for the rotated squares in the device section shown in (b) are $\theta = 29.9864^{\circ}$, $\Delta = 1.0304\ \mu\text{m}$. The lattice parameters for the rotated square lattices in the device section shown in (c) are $\theta = 25.9029^{\circ}$, $\Delta = 1.0301\ \mu\text{m}$.}
\end{figure*}

An exemplary actualized comparison of such design metrics is provided in Table \ref{tab:huangcomparison}, where we compare the design of the landmark proof-of-concept experimental device by \cite{Huang987} to device designs generated by our optimization procedures which match the lateral displacements observed by \cite{Huang987}. In this design problem, our target function represents lateral displacements at three disconnected points in radius space, and we seek to approximate the target function exclusively at those points. A device design generated with the Direct $L_2$ protocol to solve this design problem, as well as the trajectories of the particles advecting inside it, are shown in Figure \ref{fig:huangdevice2}.

The results listed in Table \ref{tab:huangcomparison} suggest that experimental realizations of device designs generated using these procedures can drastically reduce complexity and total length of such devices while matching displacement accuracy.

\begin{table}
\centering
\begin{ruledtabular}
\begin{tabular}{lccc}
\hline
& Huang (2004) & Direct $L^2$ & $\theta$-Restr. $L^2$ \\\hline
Complexity ($N$) & 8 & 2 & 2 \\
Device Length ($\ell_{\text{tot}}$) & $~$14 mm & 1.57 mm & 3 mm \\
\end{tabular}
\end{ruledtabular}
\caption{\label{tab:huangcomparison} Design metric comparison between the device described in \cite{Huang987} and device designs generated to mimic the lateral displacements observed in that device through the direct $\&$ $\theta$-restricted $L^2$ optimization processes described above. Because 3 particle size data points were surveyed in \cite{Huang987}, both $L^2$ optimization algorithms terminate at $N = 2$, as the $3$ lateral displacement functions (outlet shift + the two lattices) can fully span the space and approximate the target perfectly. Data points are approximated from \cite{Huang987} data as lateral displacements of 0.2 mm, 0.67 mm and 1.2 mm for particles of radius 0.4 $\mu$m, 0.45 $\mu$m, and 0.515 $\mu$m respectively.}
\end{table}

The device designs described in the comparison above are constructed to handle particles of only three specific sizes. However, in most colloidal suspensions encountered in nature, we find the embedded colloidal particles to possess a continuous spectrum of sizes; and designing a device with the explicit intent of precisely manipulating all of the particle sizes one can encounter in such a colloidal suspension is of paramount engineering importance. 

\begin{figure*}[h]
\centering
\includegraphics[width=1\textwidth]{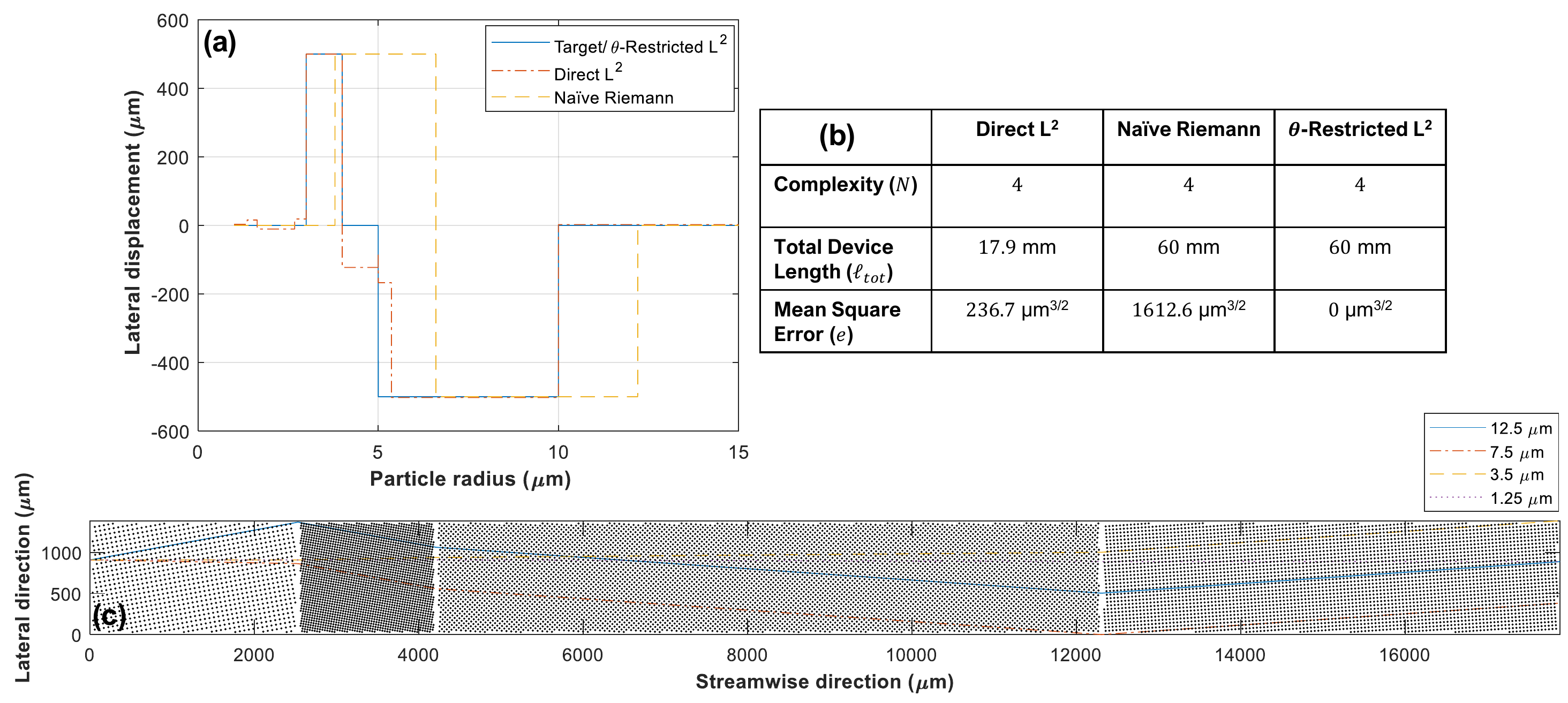}
\caption{\label{fig:bloodcomparison}Design comparisons between devices designed to separate blood components using direct $\&$ $\theta$-restricted $L^2$ optimization processes as well as the na{\"i}ve Riemann process described above. A table of design efficiency metrics for each design technique is shown in (b), their approximations to the target displacement are shown in (a), and the device design generated by the direct $L^2$ technique along with sample particle trajectories are shown in (c). The lateral displacement function domain consists of 10,000 data points distributed uniformly in the interval $[1\ \mu \text{m}, 15\ \mu \text{m}]$. Although total device length, device complexity, and lattice angles for the na{\"i}ve Riemann and $\theta$-restricted case are the same, aliasing considerably impacts the accuracy of the na{\"i}ve Riemann case. Error in the $\theta$-restricted case is within machine epsilon.}
\end{figure*}

Consequently, to determine our design protocols' efficiency at handling a continuously polydisperse suspension of particles, we test the relative merits of the design procedures listed in Section \ref{approxschemes} for a target function of continuous particle size with potential experimental use. We employ each listed procedure to generate designs for a microdevice intended to laterally displace erythrocytes and leukocytes by 500 $\mu$m in opposite directions while preventing the net lateral displacement of particulate matter not within the size ranges of either of these cell types, such as platelets. Erythrocytes are assumed to possess effective radii between 3-4 $\mu$m, while leukocytes are expected to possess radii between 5-10 $\mu$m. Efficiency metrics for designs generated using these procedures, approximations to the target lateral displacement curves, and sample particle trajectories through one of these designed devices are listed in Figure \ref{fig:bloodcomparison}.

\begin{figure*}[h]
\centering
\includegraphics[width=1\textwidth]{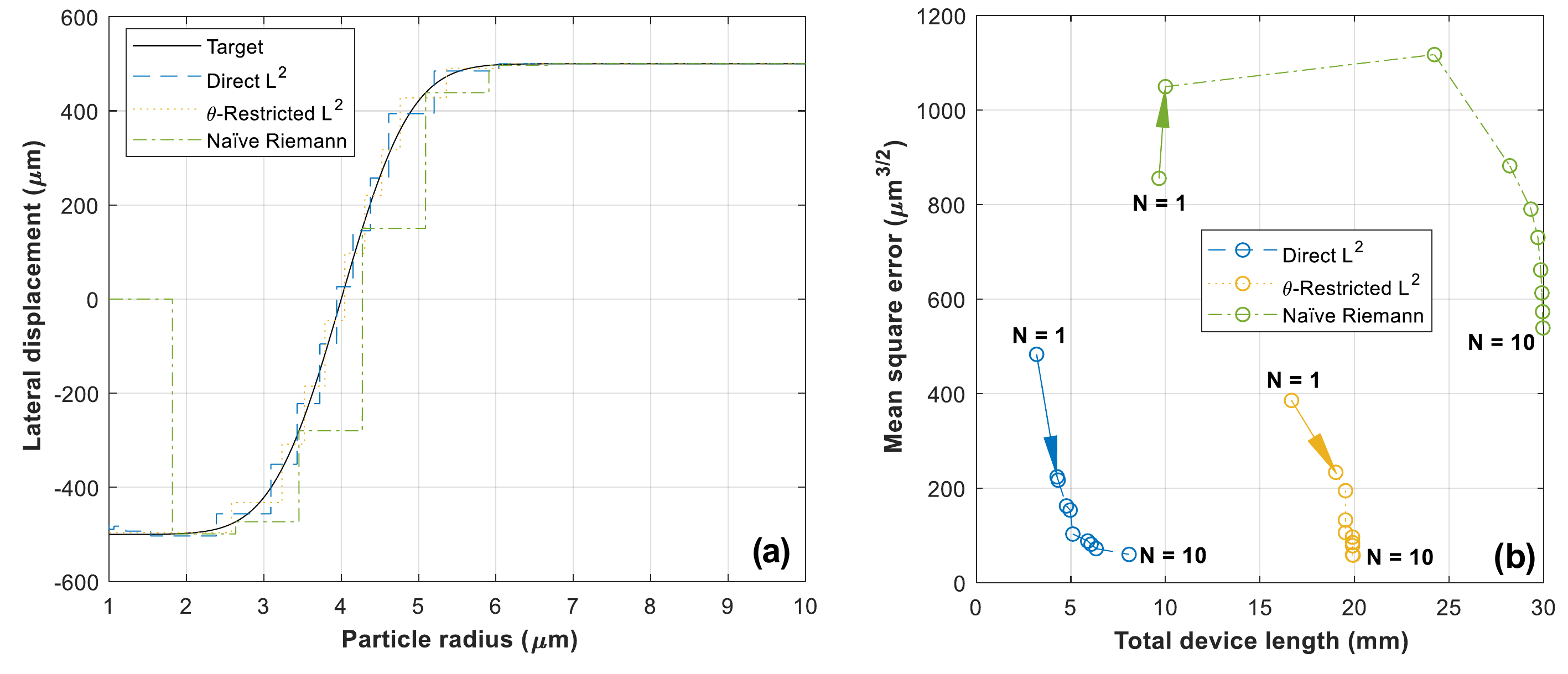}
\caption{\label{fig:sigmoid} A comparison of design efficiency metrics and lateral displacement functions for different design procedures is shown in (a), with a target displacement function $T = 500\text{erf}(r-4)$ in solid black, where both $T$ and $r$ have units of $\mu\text{m}$ and $\text{erf}(x)$ is the standard error function. The displacement function domain consists of 10,000 data points distributed uniformly in the interval $[1\ \mu \text{m}, 10\ \mu \text{m}]$. Each dot in (b) represents a device design for a given complexity $N$, while the line/arrows represent how the mean square error and total device length change as the device complexity $N$ increases. Consistent with the observations in Figure \ref{fig:bloodcomparison}, the designs generated using the direct $L^2$ procedure are far shorter and have comparable (if not smaller) mean square error for a given complexity. Also note the strictly monotonically decreasing error as a function of complexity for both $L^2$ cases, guaranteed mathematically as described in Section \ref{approxschemes}, while the na{\"i}ve Riemann case shows increasing error vs. complexity for $N < 3$.}
\end{figure*}

As a further visualization of the capabilities of these design procedures, we perform the same comparison described above, but for a continuous sigmoidal target function. For a stream of differently-sized particles entering the device at the same location, a device which sufficiently approximates this lateral displacement function would both continuously separate particles by size and cause the distance between each particle to be a Gaussian function of particle size. See Figure \ref{fig:sigmoid} for a visual representation of the sequential evolution of the design metrics as a function of $N$ for each procedure.

\section{Discussion}\label{discussion}

In this paper, we have coalesced and generalized the mathematical models for lateral displacement and collision frequency present in the literature for particles advecting through obstacle arrays of approximately infinitesimal size, utilizing a novel and comprehensive framework valid for all obstacle lattice geometries. We have then described the dependence of these key readouts as a function of both particle size and lattice parameters, and utilized this model to solve the inverse design problem of constructing a microdevice of chained obstacle lattices to fit lateral displacement or collision frequency functions of particle size.
The symmetry-induced cyclical dynamics model of particle advection through these lattices reproduces the periodic trajectories observed \& described in \cite{PhysRevLett.92.130602}\cite{PhysRevE.88.032136}\cite{PhysRevE.90.012302}, reproduces results for describing lateral displacement through square lattices in \cite{PhysRevLett.92.130602}\cite{PhysRevE.90.012302}, reproduces results for describing lateral displacement in oblique lattices in \cite{Kim201706645}, and reproduces results for collision frequencies in \cite{PhysRevE.88.032136}, all in the infinitesimal obstacle size limit.

The design algorithm described herein represents the first reported algorithmic tool for designing microdevices that sort colloidal particles by size through lateral displacement or collision frequency using microfluidic devices of chained obstacle lattices, and generates designs that are vastly improved to those in the literature with respect to microdevice size, complexity, and accuracy.

The mathematical models describing lateral displacement and collision frequency here approximate these readouts as continuous, linear functions of the streamwise length traversed by the particle which are independent of the location of initial advection of the particle relative to the lattice, whereas these are actually discontinuous functions which depend on the initial placement of the particle. For lattices that are sufficiently long in comparison to the streamwise length of the generator describing the particle's advection, the error caused by this approximation is trivial; but this source of error may become significant when a lattice's streamwise length becomes comparable to either the length of the generator for a particle advecting through that lattice or to the entrance length required for a particle to "lock in" to the mode described by its generator.

When the obstacle sizes are far smaller than the colloidal particles in these microdevices, as assumed in this work, the hydrodynamics in the device are trivial and can essentially be disregarded. When obstacles sizes do not satisfy this condition, the hydrodynamics of the system cannot be disregarded and can significantly alter the observed phenomena in comparison to our model. These sources of discrepancies include the development of lateral pressure gradients as described by \cite{C7LC00785J}, edge effects at the interface between different chained lattices, and differences in particle advection trajectories post-collision due to the now non-trivial flow perturbation induced by the obstacles. However, work by Risbud \& Drazer\cite{risbud_drazer_2013}\cite{PhysRevE.90.012302} has identified that in systems with arbitrary obstacle sizes relative to the colloidal particles, the interaction dynamics between particles and obstacles are essentially identical to those described in Section \ref{traj} except that the "output" displacements are $\pm b_c$, a parameter depending on the hydrodynamics of the microdevice and the particle/obstacle sizes, rather than $\pm r$. Since the key features of the symmetry-induced cyclical dynamics in this system are independent of the magnitude of the "output" displacement from a single collision, one can approximate the lateral displacement and collision frequency functions of size in systems with nontrivial obstacle sizes by re-scaling the displacements from $r$ to $b_c$ and using the exact same mathematical techniques we detail above. This approximation is supported by noting that the Risbud--Drazer parameter $b_c$ converges to $r$ as the obstacle radius approaches 0 in their model.

In our model, we also neglect other transport effects, such as diffusion and particle-particle interaction. These phenomena are neglected due to the strong, dynamically limiting nature of the symmetry-induced cyclical dynamics in these devices; unless a particle's size is very near to a critical radius, its trajectory will be stable to lateral motion perturbations, and will "reset" upon collision with an obstacle, stabilizing the trajectories further and causing the effective Peclét number of the system to increase dramatically in comparison to an unpatterned microchannel of equivalent size. The effects of these phenomena should then make themselves clear in regions where the lateral displacement and collision frequencies possess discontinuities as a function of size, where these discontinuities will appear mollified and sigmoidal in agreement with observations from \cite{PhysRevE.88.032136}.

The phenomenon responsible for the dominant physics in the system, which we term symmetry-induced cyclical dynamics, is independent of the nature of the particle-obstacle interaction as long as the interaction produces a discrete set of possible lateral displacements. As a result, particle sorting microdevices where the particle-obstacle interaction is mediated not just by the contact force between particles and obstacles, but by other forces (such as electromagnetic forces) that depend on properties of the colloidal particles which are independent of its size, have been constructed successfully.\cite{C8LC01416G} Utilizing the design protocol described here for designing thse types of devices is particularly promising as studies indicate that cancer cells with particularly malignant or drug-resistant attributes have differing electromagnetic properties while having approximately the same size.\cite{8631175}
\section{Summary \& Conclusions}

We have unified and expanded the different mathematical models in the literature describing lateral displacement and collisions of particles advecting through specific lattices of infinitesimal obstacles into a general mathematical framework, and then used this framework to obtain key size-dependent transport quantities such as lateral displacement per length and spatial collision frequency for arbitrary infinitesimal obstacle lattices. We then utilized the predictions for lateral displacement functions from this framework to show that, by chaining rotated square infinitesimal obstacle lattices together, we can recreate essentially any lateral displacement function of particle size to arbitrary accuracy. Afterwards, we described different microdevice design protocols to approximate target lateral displacement functions, and demonstrated their effectiveness by generating designs approximating lateral displacement functions both in the literature and with key potential applications.

\bibliography{references}

\end{document}